\documentclass[aps,a4paper,superscriptaddress]{revtex4}
\usepackage{tensor}
\usepackage{graphicx}
\usepackage{amsmath}
\usepackage{amssymb}
\usepackage{enumerate}
\usepackage{subfigure}
\usepackage{tabularx}
\usepackage[colorlinks=true, pdfstartview=FitV, linkcolor=blue, citecolor=red, urlcolor=black, breaklinks=true]{hyperref}
\newcommand{\be}{\begin{equation}}
\newcommand{\ee}{\end{equation}}
\newcommand{\ben}{\begin{eqnarray}}
\newcommand{\een}{\end{eqnarray}}
\newcommand{\bes}{\begin{subequations}}
\newcommand{\ees}{\end{subequations}}
\def\bal#1\eal{\begin{align}#1\end{align}}

\newcommand{\sech}{{\rm sech}}
\newcommand{\LL}{{\mathcal L}}

\newcommand{\Hc}{\mathcal{H}}
\begin{document}
\title{Novel way to construct spatially localized finite energy structures}
\affiliation{Departamento de F\'\i sica, Universidade Federal da Para\'\i ba, 58051-970 Jo\~ao Pessoa, PB, Brazil}
\author{D. Bazeia}\email{dbazeia@gmail.com}
\affiliation{Departamento de F\'\i sica, Universidade Federal da Para\'\i ba, 58051-970 Jo\~ao Pessoa, PB, Brazil}
\author{M.A. Marques}\email{marques@cbiotec.ufpb.br}
\affiliation{Departamento de Biotecnologia, Universidade Federal da Para\'\i ba, 58051-900 Jo\~ao Pessoa, PB, Brazil}
\affiliation{Departamento de F\'\i sica, Universidade Federal da Para\'\i ba, 58051-970 Jo\~ao Pessoa, PB, Brazil}
\author{M. Paganelly}\email{matheuspaganelly@gmail.com}
\affiliation{Departamento de F\'\i sica, Universidade Federal da Para\'\i ba, 58051-970 Jo\~ao Pessoa, PB, Brazil}
\begin{abstract}
In this work we introduce a procedure to find localized structures with finite energy. We start dealing with global monopoles, and add a new contribution to the potential of the scalar fields, to balance the contribution of the angular gradients of the fields which lead to a slow falloff in the energy density. Within the first order formalism, first order equations that are compatible with the equations of motion are obtained and the stability under small fluctuations is investigated. We then include another set of scalar fields and study how it contributes to change the profile of the localized structure. We also study how these configurations modify the electric properties of a system with a single point charge, with generalized electric permittivity controlled by scalar fields. In this new model, in particular, we show that, depending on the specific modification of the electric properties of the medium, the electric field may engender the unusual behavior of pointing towards a positive charge.
\end{abstract}
\maketitle

\section{Introduction}
In accordance with the Big Bang theory, the Universe has evolved until the present days expanding and decreasing its density and temperature, resulting in several phase transitions in the early times of the cosmic history. Associated with the mechanism of spontaneous symmetry breaking, some phase transitions may have given rise to topological structures \cite{kibble} that may have contributed to the formation of structures in the Universe. Several of these elementary topological structures appear in high energy physics under the action of scalar and gauge fields, and the most known ones are kinks, vortices and monopoles \cite{manton,vilenkin}. Monopoles, in particular, arise in three spatial dimensions. To investigate them, one in general consider a triplet of scalar fields whose associated Lagrange density engenders a global symmetry with a dynamical term and a potential one. However, the model leads to infinite energy \cite{vilenkinprl}. To regularize its energy, one may consider a model with $SU(2)$ local symmetry that couples the aforementioned scalar fields to gauge fields \cite{thooft,polyakov}. By doing so, one gets finite energy configurations that supports a first order framework \cite{ps,bogo}, that is, one may find first order differential equations that support minimum energy solutions which solve the corresponding equations of motion. These solutions are usually referred to as Bogomol'nyi-Prasad-Sommerfield (BPS) states.

Even though global monopoles have divergent energy due to the slow falloff of the contribution of the angular gradients of the fields, their stability has been the subject of study over the years in the literature \cite{stab1,stab2,stab3,stab4}. In Ref.~\cite{stab4}, in particular, the authors showed that spherically symmetric solutions are classically stable to axially symmetric, square integrable or power-law decay fluctuations. Moreover, although global monopoles have infinite energy in flat spacetime, they find several applications in the curved spacetime; see Refs.~\cite{vilenkinprl,g1,g2,g3,g4,g5,g51,g6,g61,g7}. In particular, in Ref.~\cite{vilenkinprl} the authors studied the Einstein equations for the metric outside a global monopole and found that the monopole does not make any gravitational force on the nonrelativistic matter, but the space around it has a deficit solid angle and light rays are all deflected by the same angle. In Ref.~\cite{g1}, the cosmological evolution of global monopoles was investigated and the authors found that, in specific conditions, these structures may become seeds for galaxy and large-scale structure formation. More recently, in \cite{g5} the authors studied a class of models of topological inflation in which global monopole may seed inflation and help 
providing solution to the graceful exit problem of topological inflation. Also, in \cite{g51} the investigation revealed that after a transition in which global monopoles form, spatial sections of a spatially flat and infinite Universe becomes finite and closed. Moreover, in Ref.~\cite{g6}, the authors found magnetic monopoles induced by global monopoles in the presence of a Kalb-Ramond axion field. Global monopoles were also investigated in the context of Palatini gravity, in which the behavior of the solution depends on the parameters in the model, presenting similar characteristics of the Schwarzschild or the Reissner-Nordstr\"om metric, both with a monopole charge in Einstein's General Relativity \cite{g7}. Yet in \cite{g61} the presence of rotating global monopole was considered to contribute to modify the deflection angle of light for an observer and source at finite distance from the rotating global monopole. 

Motivated by the above recent results, in which one uses global monopole to describe problems of current interest in high energy physics, in this work we return to the problem of global monopole with divergent energy in flat spacetime. In Sec. \ref{monopole}, in particular, we focus on modifying the model to describe a procedure that allows the presence of global monopoles with finite energy. We also study stability, and show that they are linearly stable, with the stability equation written in terms of first order differential operators. These new results suggested that we enhance the symmetry of the model, adding other degrees of freedom and finding exact solutions in this much more intricate situation, as described in Sec. \ref{extended}. Here, in particular, we study two distinct models and show how the new fields contribute to modify the profile of the localized structure in a significant manner. Inspired by these results and by the recent work on electrically charged localized structures \cite{elect}, in Sec. \ref{electric} we also investigate the behavior of a single point electric charge in a medium where the electric permittivity is driven by scalar fields. We end the work in Sec. \ref{end}, adding some comments and conclusions, and describing new possible lines of future research related to the main results of the present work. 

\section{Global Monopole}
\label{monopole}

We start our investigation with the Lagrange density below in $(D,1)$ flat spacetime dimensions, with $D\geq3$. Although we are interested in the case $D=3$, the investigation is carried out in $D\geq 3$ with no additional difficulties. Here the  metric tensor is diagonal, with components diag$(\eta_{\mu\nu})=(-,+,+,+,\cdots)$, and we have
\be\label{model}
\LL_1 = -\frac12 \partial_\mu\phi^a\partial^\mu\phi^a-V(|\phi|),
\ee
where $|\phi|= \sqrt{\phi^a\phi^a}$, with $a=1,\ldots,D$. By varying the action associated to it, we get the equation of motion
\be\label{eomt}
\partial_\mu\partial^\mu \phi^a = \frac{\phi^a}{|\phi|}V_{|\phi|},
\ee
where $V_{|\phi|}=dV/d|\phi|$. To study global monopoles, we consider static configurations with the ansatz
\be
\label{ansatz}
\phi^a = \frac{x_a}{r} H(r),
\ee
where $|\phi| = H(r)$. The equation of motion \eqref{eomt} with the above ansatz take the form
\be\label{eom}
\frac{1}{r^{D-1}}\left(r^{D-1} H^\prime \right)^\prime = \left(D-1\right)\frac{H}{r^2} + V_{H}.
\ee
The energy density is calculated standardly; it is given by
\be\label{eden}
\rho = \frac12{H^\prime}^2 + \frac{D-1}{2}\frac{H^2}{r^2} + V(H).
\ee
Unfortunately, as it is known, topological solutions in this model leads us to infinite energy. Indeed, we can define $r_{core}$, that determines the size of the structure core, such that for $r>>r_{core}$, we have $H\approx \eta$, $H^\prime\approx0$ and $V\approx0$. In this regime, the energy density behaves as $\rho\propto 1/r^2$, which leads to infinite energy; see Ref.~\cite{vilenkinprl}.

In order to circumvent the above issue, we follow the lines of Ref.~\cite{bogo,prl}. We then introduce an auxiliary function $W=W(|\phi|)$ and rewrite the energy density in the form
\be
\begin{aligned}
\label{bogomolnyi}
	\rho = \frac12\left(H^\prime \mp \frac{W_{H}}{r^{D-1}} \right)^2 + V(H) + \frac{D-1}{2}\frac{H^2}{r^2}-\frac12\frac{W_{H}^2}{r^{2D-2}}\pm \frac{1}{r^{D-1}}W^\prime.
\end{aligned}
\ee
We now consider that the potential also depends explicitly on the radial coordinate, so we write \cite{prl}
\be
\label{potential}
V(r,|\phi|) = \frac{W_{|\phi|}^2}{2r^{2D-2}}-\frac{D-1}{2}\frac{|\phi|^2}{r^2}.
\ee
In this situation, the energy density becomes
\be
\label{energydensity}
\rho = \frac12\left(H^\prime \mp \frac{W_{H}}{r^{D-1}}\right)^2 \pm \frac{1}{r^{D-1}}W^\prime,
\ee
such that the energy is bounded
\be
\label{EB}
E\geq E_B = \Omega_{(D)}\,|W(H(\infty)) - W(H(0))|,
\ee
where $\Omega_{(D)} = 2\pi^{D/2}/\Gamma(D/2)$ denotes the $D$-dimensional solid angle contribution. The energy is then minimized to $E=E_B$ if the solutions obey the first order equations
\be
\label{firstorder}
H^\prime =\pm \frac{W_{H}}{r^{D-1}}.
\ee
One can show this first order equation is compatible with the equation of motion \eqref{eom}. To understand this equation better, one can make the change of variable $dx=\pm dr/r^{D-1}$, which leads us to the correspondence $x=\mp 1/((D-2)r^{D-2})$, to show that the above expression can be rewritten in the form $dH/dx=W_H$; see \cite{prl} for further details. The latter equation arises in the study of kink configurations in $(1,1)$ spacetime dimensions \cite{vachaspati}. In this sense, the aforementioned change of variable maps the one-dimensional kink solution into the global monopole that we are searching for, but we have to be careful, since the upper sign maps the interval $x\leq0$ and the lower sign is associated to the interval $x\geq0$ of the kink solution; see Ref.~\cite{prl}.

Considering the potential in Eq.~\eqref{potential}, one can rewrite Eq.~\eqref{eom} in the form $r^{D-1}\left(r^{D-1} H^\prime \right)^\prime = U_{H}$ to see that
\be\label{eff}
U(|\phi|)= \frac12 W_{|\phi|}^2
\ee
plays the role of an effective potential in the model. Indeed, the minima of this potential are connected by the one-dimensional kink solutions that are related with the monopole ones by the change of variables discussed below Eq.~\eqref{firstorder}. In the global monopole scenario, the energy density can be written in terms of $U(|\phi|)$ as
\be
\rho = \frac12{H^\prime}^2 + \frac{1}{r^{2D-2}}\,U(H).
\ee
Thus, the above formalism allows us to find global monopoles with finite energy, an unprecedented achievement in the related literature. As we can see from the above calculation, to get rid of the portion of energy density that appears as the second term in Eq. \eqref{eden}, we had to make the potential explicitly dependent of the radial coordinate that describes the global monopole. The procedure breaks translational invariance, but this is also present in several distinct situations of physics interest, for instance, in lattice systems in condensed matter, where the periodicity of the lattice may change the spectrum of the system into a band-like structure which is of fundamental importance to the understanding of the electric conduction in metals, to quote just one possibility. In a way similar to this, the breaking of translational invariance has also been used in Ref. \cite{HL} to study optical conductivity by means of gauge/gravity duality which relates a theory of gravity to a non-gravitational theory that lives in a lower dimensional space. The addition of holographic lattices with the inclusion of scalar fields to simulate a lattice unveils an interesting scenario to make the conductivity finite. Another procedure to holography without translational symmetry can be found in \cite{HWT}, where the author suggests a conceptually distinct approach as an effective bulk description of a theory without momentum conservation, via massive gravity. The breaking may also appear spontaneously, as it was shown sometime ago in Ref. \cite{SM} in a simple model composed of two distinct charge carries and vacancies on a ring, in which the time evolution under specific rules may lead to spatial particle condensation and the corresponding breaking of translational invariance. There are several other issues related to the breaking of translational invariance, some of them have been considered for vortices in the presence of electric and magnetic impurities in the plane \cite{HKT,Tong,VI} and for kinks also in the presence of impurity in the real line \cite{Adam,Manton,K1}, in direct connection with the presence of localized configurations that preserve the BPS structure of the model. 

These recent investigations motivate us to further study the robustness of the global monopoles on general grounds. From the mathematical computation described above, we notice that, even though the effective potential \eqref{eff} is non-negative, there is a negative term in the potential \eqref{potential} that defines the model \eqref{model}. This suggests that we investigate the stability of the solutions of Eq.~\eqref{firstorder} under small fluctuations, to get more information on the robustness of the localized structures. To do this, we consider the scalar fields in the form
\be\
\phi^{a}(r,t)=\phi^{a}(r)+\eta^{a}(r,t),
\ee
where the first term is the static solution, governed by Eq.~\eqref{eom}, and the second one is a small time-dependent perturbation. Using the above expression in the time-dependent equation of motion \eqref{eomt}, we obtain
\be
\label{secondorder}
\partial_{\mu}\partial^{\mu}\eta^{a}-V_{|\phi||\phi|}\eta^{a}=0.
\ee
Since the static configurations support the symmetry in Eq.~\eqref{ansatz}, we consider the fluctuations to have a similar form, given by
\be\label{fluc}
\eta^{a}(r,t)=\frac{x^{a}}{r}\sum_{k}\xi_{k}(r)\cos{(\omega_{k}t)}.
\ee
In this case, the stability equation \eqref{secondorder} becomes
\be
-\frac{1}{r^{D-1}}(r^{D-1}\xi'_{k})'+\left(\frac{(D-1)}{r^{2}}+V_{|\phi||\phi|}\right)\xi_{k}=\omega^{2}_{k}\xi_{k}.
\ee
This expression is valid for any solution of the equation of motion \eqref{eom}. We now consider the static configurations to be described by the first order framework that we developed in this paper. So, we take the potential $V(|\phi|)$ in the form of Eq.~\eqref{potential}. By doing this, the above equation turns into
\be\label{stab}
-\frac{1}{r^{D-1}}(r^{D-1}\xi'_{k})'+\frac{W_{H}W_{HHH} + W^{2}_{HH}}{r^{2D-2}}\xi_{k}=\omega^{2}_{k}\xi_{k},
\ee
which governs the behavior of the fluctuations \eqref{fluc}. The solution $H(r)$ is stable if $\omega^{2}_{k}\geq0$. One can rewrite the above equation in terms of a second order operator, $L$, as $L\xi_{k}=\omega^{2}_{k}\xi_{k}$. One can show that the operator $L$ may be written in the form $L=S^{\dagger}S$, where
\bes
\bal
S &=-\frac{d}{dr}+\frac{W_{HH}}{r^{D-1}},\\
S^\dagger &=\frac{d}{dr}+\frac{W_{HH}}{r^{D-1}} + \frac{(D-1)}{r}.
\eal
\ees
The factorization of the operator $L$ in terms of the supersymmetric operators $S$ and $S^{\dagger}$ ensures that the eigenvalue equation \eqref{stab} only supports non negative eigenvalues, i.e., $\omega^{2}_{k}\geq 0$. Thus, the solutions of Eq.~\eqref{firstorder} are stable under small fluctuations, and this ends the general investigation.

		\begin{figure}[t!]
		\centering
		\includegraphics[width=6.2cm,trim={0cm 0cm 0 0},clip]{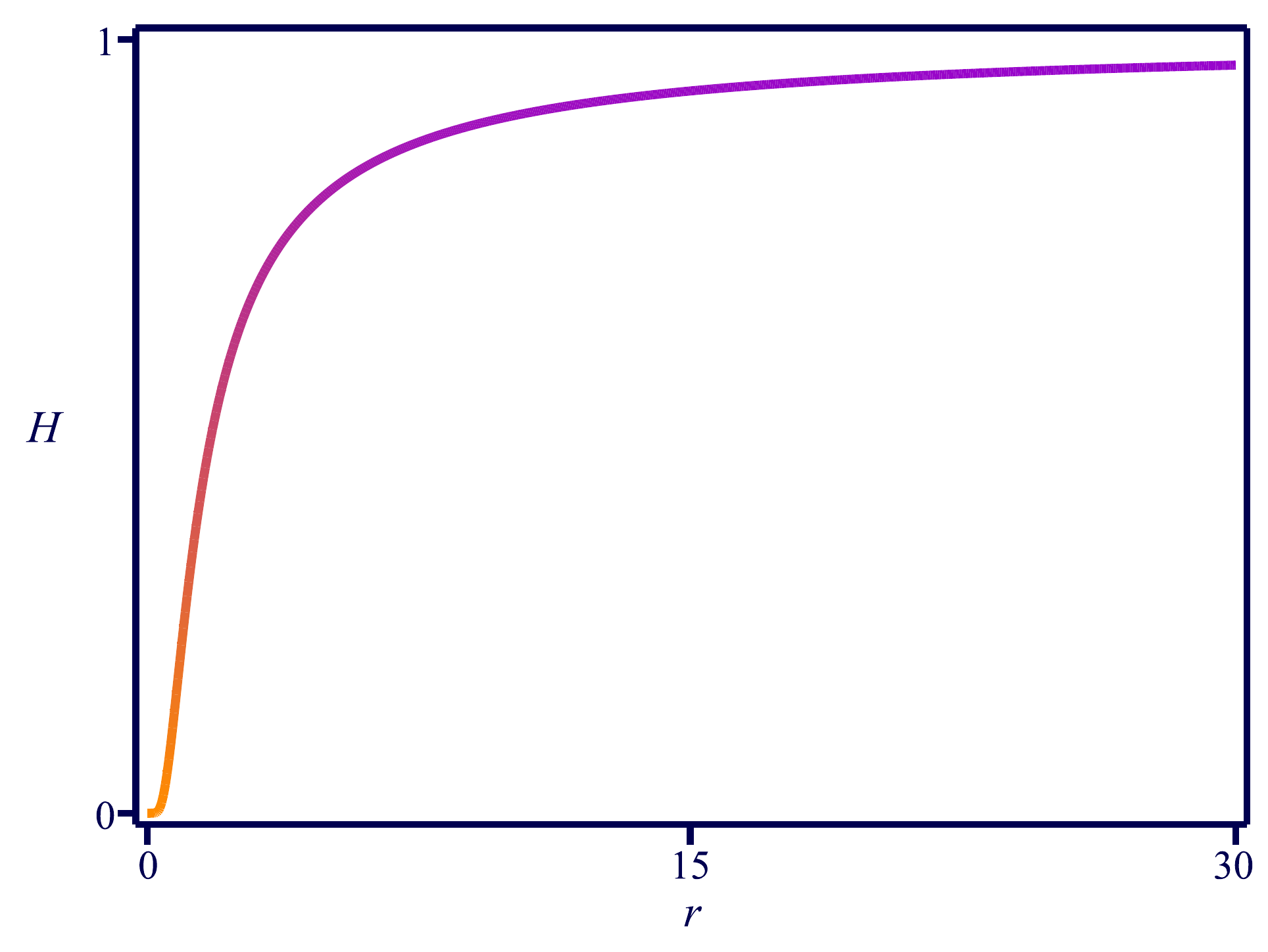}
		\includegraphics[width=6.2cm,trim={0cm 0cm 0 0},clip]{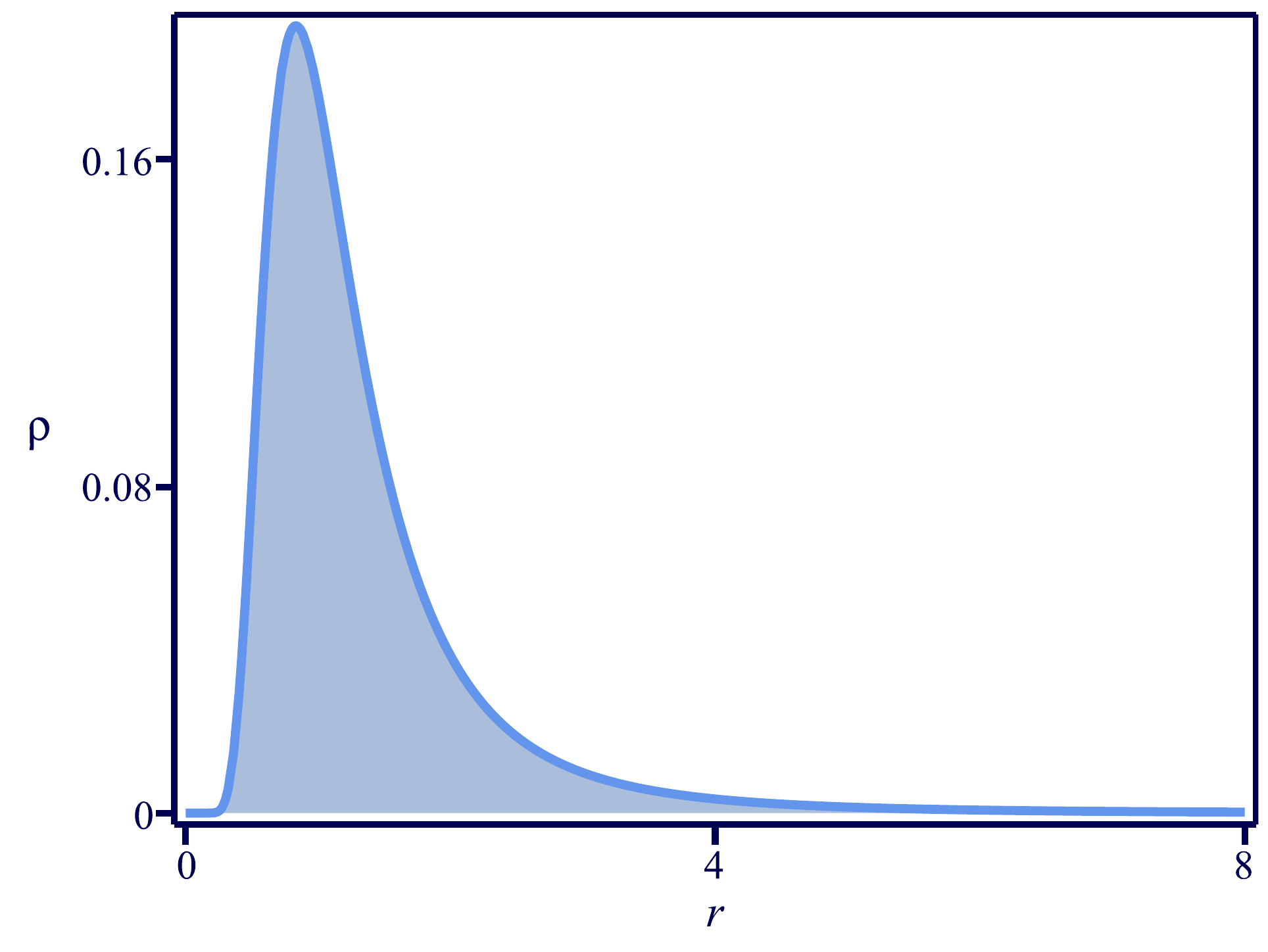}
		\includegraphics[width=6.2cm,trim={0cm 0cm 0 0},clip]{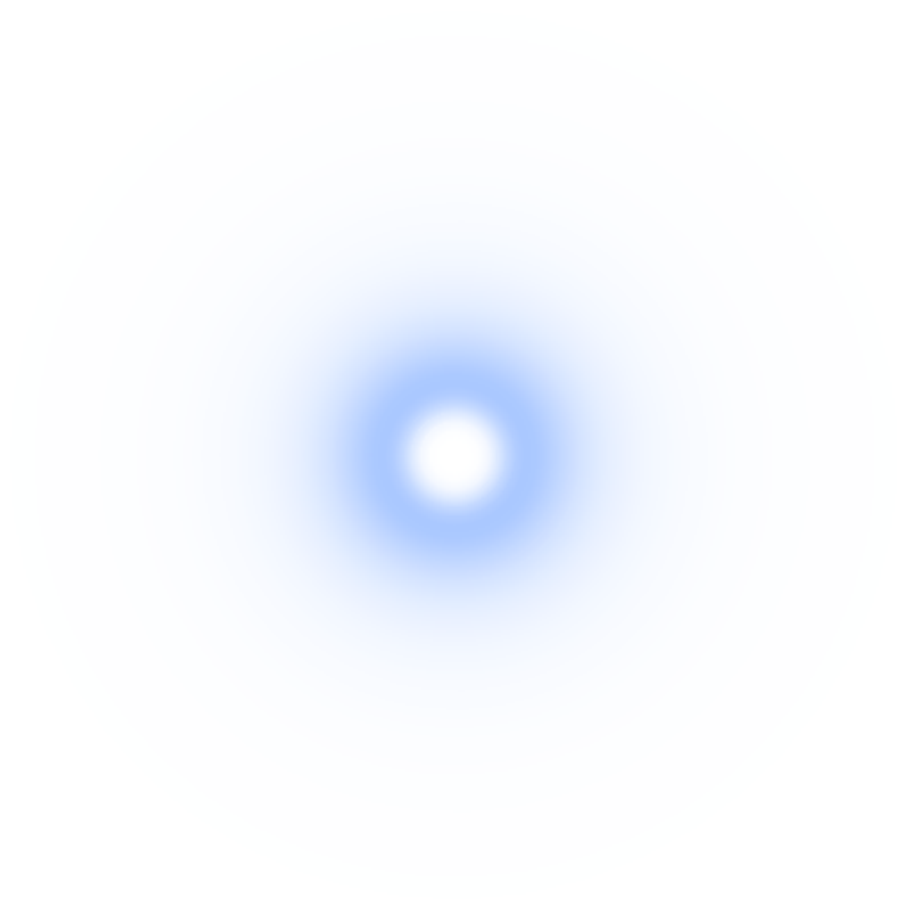}
		\caption{The solution in Eq.~\eqref{sol1} (top left), the energy density \eqref{rho1} (top right), and the planar section passing through the center of the energy density (bottom), for $D=3$.}
		\label{fig1}
		\end{figure}

\subsection{The $|\phi|^4$ model}
\label{monopolea}

To illustrate the above procedure, let us present a model described by the function $W(|\phi|)$ associated to a $|\phi|^4$ model; it is given by
\be
\label{p1}
W(|\phi|)=|\phi|^2-\frac{1}{3}|\phi|^{3}.
\ee
The effective potential in Eq.~\eqref{eff} takes the form
\be
U(|\phi|) = \frac12|\phi|^2\left(2-|\phi|\right)^2.
\ee
Notice that its minima are located at $|\phi|=0$ and $|\phi|=2$. Also, there is a local maximum at $|\phi|=1$, such that $U(1)=1/2$. The first-order equation \eqref{firstorder} with positive sign reads
\be\label{fophi4}
H^\prime = \frac{H(2-H)}{r^{D-1}}.
\ee
It supports the solution
\be\label{sol1}
H(r)=1-\tanh\bigg(\frac{1}{(D-2)\,r^{D-2}}\bigg). 
\ee
This solution connects the points $\phi=0$ and $\phi=1$. The energy density \eqref{energydensity} becomes
\be\label{rho1}
\rho = \frac{1}{r^{2D-2}}\sech^4\bigg(\frac{1}{(D-2)\,r^{D-2}}\bigg).
\ee
Integrating the energy density we obtain the energy $E=2\,\Omega_{(D)}/3$, matching with the minimum energy in Eq.~\eqref{EB} for the function $W$ in Eq.~\eqref{p1}. We plot the solution in Eq.~\eqref{sol1} and the above energy density in Fig.~\ref{fig1} for $D=3$. We see from the energy density that the structure presents a hole at its center, so it is a hollow global monopole, similar to the magnetic monopole recently described in \cite{olmo}. We also note that its maximum is around $0.20$.

\subsection{The $|\phi|^6$ model}
\label{monopoleb}

\begin{figure}[t!]
		\centering
		\includegraphics[width=6.2cm,trim={0cm 0cm 0 0},clip]{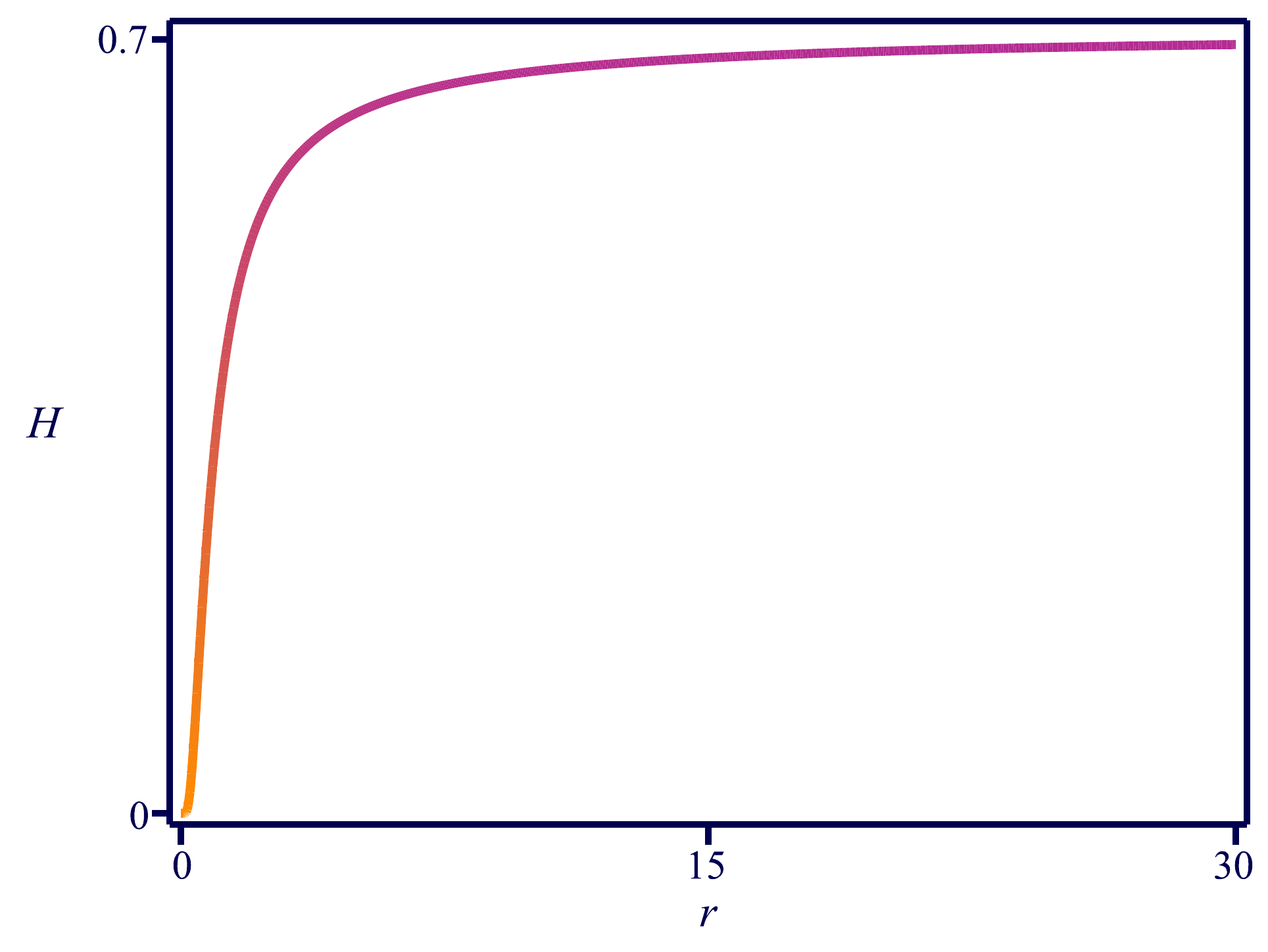}
		\includegraphics[width=6.2cm,trim={0cm 0cm 0 0},clip]{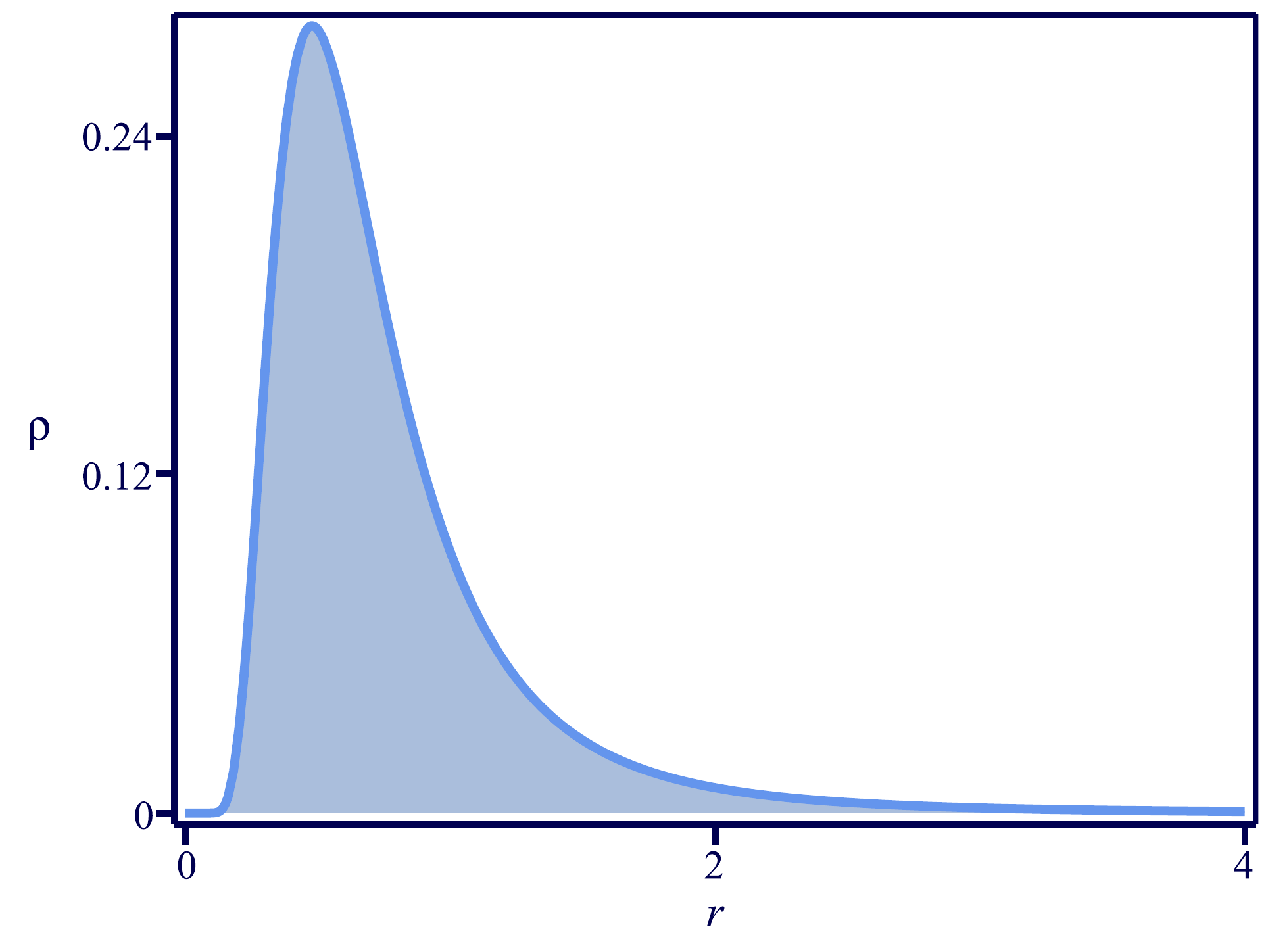}
		\includegraphics[width=6.2cm,trim={0cm 0cm 0 0},clip]{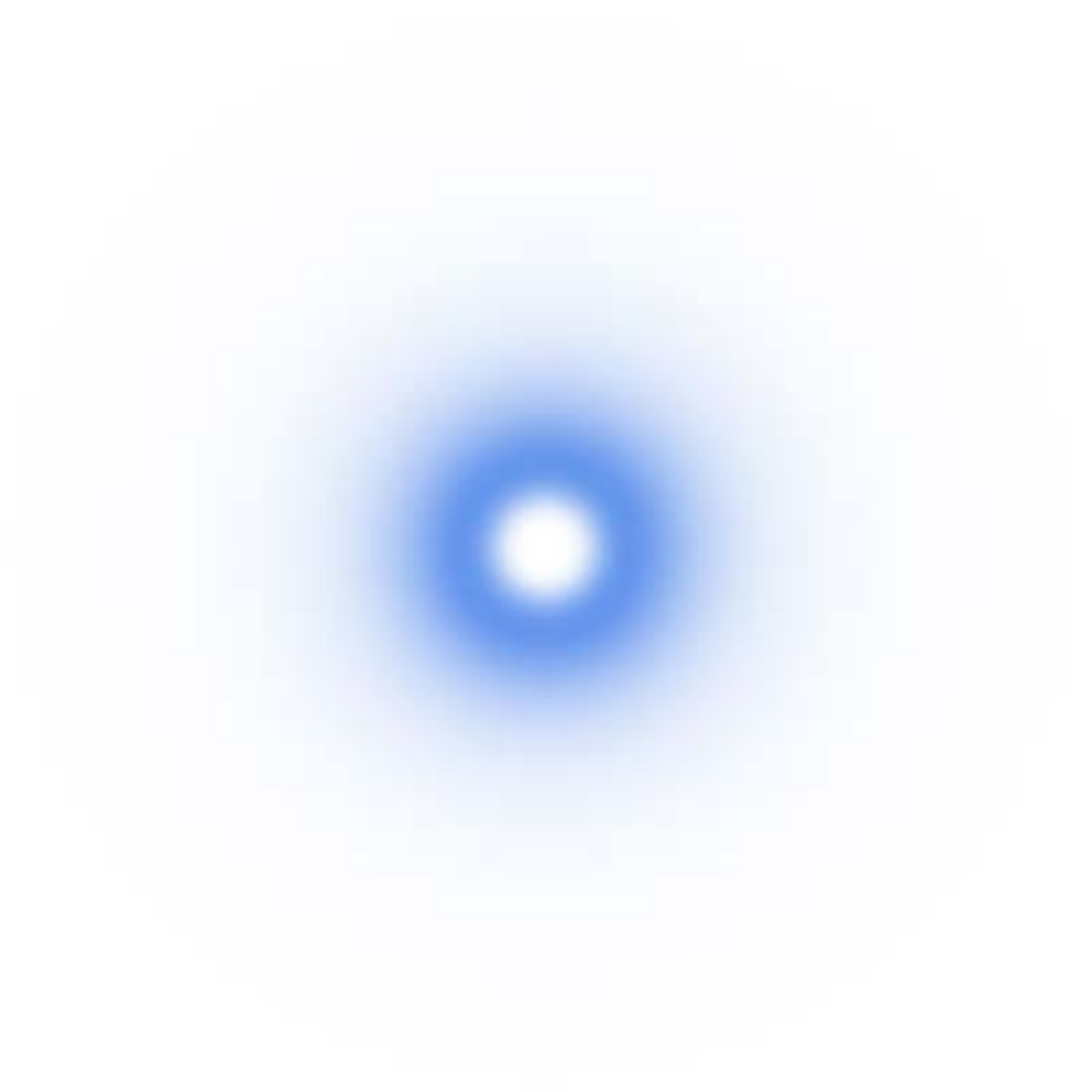}
		\caption{The solution in Eq.~\eqref{solphi6} (top left), the energy density \eqref{rhophi6} (top right) and the planar section passing through the center of the energy density (bottom), for $D=3$.}
		\label{fig2}
		\end{figure}

As a second example, we consider the function
\be\label{p2}
W(|\phi|) = \frac12|\phi|^2 - \frac14|\phi|^4.
\ee
From Eq.~\eqref{eff} we get a $|\phi|^6$ effective potential, in the form
\be
U(|\phi|) = \frac12|\phi|^2\left(1-|\phi|^2\right)^2.
\ee
It supports minima at $|\phi|=0$ and $|\phi|=1$, and a set of local maxima at $|\phi|=1/\sqrt{3}$, such that $U(1/\sqrt{3})=2/27$. The first order equation \eqref{firstorder} with the positive sign becomes
\be
H^\prime = \frac{H\left(1-H^2\right)}{r^{D-1}}.
\ee
One can show that it supports the solution
\be\label{solphi6}
H(r) = \sqrt{\frac12 -\frac12\tanh\left(\frac{1}{(D-2)\,r^{D-2}}\right)}.
\ee
This solution connects the points $\phi=0$ and $\phi=1/\sqrt{2}$, as expected from the correspondence discussed below Eq.~\eqref{firstorder}. In this case, the energy density in Eq.~\eqref{energydensity} reads
\be\label{rhophi6}
\begin{aligned}
	\rho(r) = \frac{1}{8r^{2D-2}}\,\left(1+\tanh\left(\frac{1}{(D-2)\,r^{D-2}}\right)\right)
			\;\sech^2\left(\frac{1}{(D-2)\,r^{D-2}}\right).
\end{aligned}
\ee
In Fig.~\ref{fig2}, we depict the solution \eqref{solphi6} and the above energy density for $D=3$. By integrating the above expression, one gets the energy $3\Omega_{(D)}/16$, in accordance with the Bogomol'nyi bound in Eq.~\eqref{EB}. It engender a hole at its center, also being a hollow global monopole. And also, its maximum is below $0.30$, a little higher than it is in the previous model, but its hollow portion is smaller, making the blue color in the planar section of its energy density darker, as we see when comparing the two Figs. \ref{fig1}  and \ref{fig2}. 
\section{Extended global monopole}
\label{extended}

Motivated by the results obtained with the above $|\phi|^4$ and $|\phi|^6$ models, let us now study two other models, extended to accommodate other sets of scalar fields.

\subsection{First extended model}

Let us now consider the extended model
\be\label{model2}
\LL_2= -\frac12 \partial_\mu\phi^a\partial^\mu\phi^a -\frac12 \partial_\mu\chi^a\partial^\mu\chi^a-V(|\phi|,|\chi|),
\ee
where the additional family of fields $\chi^a$ is also such that $|\chi|= \sqrt{\chi^a\chi^a}$, with $a=1,\ldots,D$ and $D\geq 3$. The coupling between the two family of fields is in the potential. In the context where one family is in the visible sector, and the other in the hidden or dark sector this is usually referred to as the Higgs portal; see, e.g., the recent review on the subject of dark matter through the Higgs portal \cite{HP}. Of course, Ref. \cite{HP} focuses mainly on dark matter phenomenology, so it may serve here to motivate the construction of models that support topological structures such as monopoles. In this sense, the above model may be seem as an extension of the model described in Sec. \ref{monopole} to the case where two distinct global monopoles may communicate with each other through the Higgs portal in an analytical way, stimulating new investigations on the subject. 

We then study the presence of localized structure. One first notices that the above model describes two families of equations of motion, which can be written in the form
\bes\label{eomt2}
\bal
\partial_\mu\partial^\mu \phi^a = \frac{\phi^a}{|\phi|}V_{|\phi|},\\
\partial_\mu\partial^\mu \chi^a = \frac{\chi^a}{|\chi|}V_{|\chi|},
\eal
\ees
where $V_{|\phi|}=\partial V/\partial|\phi|$ and $V_{|\chi|}=\partial V/\partial|\chi|$. In order to search for field configurations with the monopole profile, we use the ansatz in Eq.~\eqref{ansatz} and also
\be
\label{ansatzchi}
\chi^a = \frac{x_a}{r} \Hc (r).
\ee
The equations of motion \eqref{eomt2} are then reduced to  
\bes\label{eom2}
\bal
\frac{1}{r^{D-1}}\left(r^{D-1} H^\prime \right)^\prime = \left(D-1\right)\frac{H}{r^2} + V_{H},\\
\frac{1}{r^{D-1}}\left(r^{D-1} \Hc^\prime \right)^\prime = \left(D-1\right)\frac{\Hc}{r^2} + V_{\Hc},
\eal
\ees
and the energy density becomes
\be
\rho = \frac12{H^\prime}^2 + \frac12{\Hc^\prime}^2 + \frac{D-1}{2r^2}\left(H^2+\Hc^2\right) + V(H,\Hc).
\ee
In a way similar to the previous case, here we introduce the function $W=W(|\phi|,|\chi|)$ to write
\be
\begin{aligned}
\label{bogomolnyi2}
	\rho = \frac12\left(H^\prime \mp \frac{W_{H}}{r^{D-1}} \right)^2\! + \frac12\left(\Hc^\prime \mp \frac{W_{\Hc}}{r^{D-1}} \right)^2\!+ V(H,\Hc) + \frac{D-1}{2r^2}\left(H^2+\Hc^2\right)
-\frac{1}{2r^{2D-2}}\left(W_{H}^2 + W_{\Hc}^2\right)\pm \frac{1}{r^{D-1}}W^\prime.
\end{aligned}
\ee

In this case, if the potential is
\be
\begin{aligned}
	\label{potential2}
V(r,|\phi|,|\chi|) = \frac{1}{2r^{2D-2}}\left(W_{|\phi|}^2 + W_{|\chi|}^2\right)
-\frac{D-1}{2r^2}\left(|\phi|^2+|\chi|^2\right),
\end{aligned}
\ee
the energy density becomes
\be
\label{energydensity2}
\rho = \frac12\left(H^\prime \mp \frac{W_{H}}{r^{D-1}}\right)^2 + \frac12\left(\Hc^\prime \mp \frac{W_{\Hc}}{r^{D-1}}\right)^2 \pm \frac{1}{r^{D-1}}W^\prime.
\ee
The energy can then be bounded
\be
\label{EB2}
E\geq E_B = \Omega_{(D)}\,|W(H(\infty),\Hc(\infty)) - W(H(0),\Hc(0))|,
\ee
getting its minimum value, $E=E_B$, for solutions that obey the first order equations
\be
\label{firstorder2}
H^\prime =\pm \frac{W_{H}}{r^{D-1}} \quad\text{and}\quad \Hc^\prime = \pm \frac{W_{\Hc}}{r^{D-1}}.
\ee
An interesting model, which allows the presence of analytical solutions is given by
\be
W(|\phi|,|\chi|) = |\phi|^2-\frac13|\phi|^3-s\left(|\phi|-1\right)|\chi|^2.
\ee
where $s$ is a real parameter that controls the coupling between the two sets of fields, that is, it controls the Higgs portal. In this case the first order equations can be written as 
\be\label{foeq}
H^\prime = \frac{H(2-H)-s\Hc^2}{r^{D-1}}\quad\text{and}\quad \Hc^\prime = -\frac{2s(H-1)\Hc}{r^{D-1}}.
\ee
These equations admit analytical solutions for different values of $s$, and for $s=0$ we get back to the previous model described by the first order Eq. \eqref{fophi4}. An interesting set of analytical solutions is obtained for $s\in(0,1/2)$. It is given by
\bes\label{solbnrt}
\bal
H(r) &= 1-\tanh\left(\frac{2s}{(D-2)r^{D-2}}\right),\\
\Hc(r)&=\sqrt{\frac{1-2s}{s}}\;\sech\left(\frac{2s}{(D-2)r^{D-2}}\right).
\eal
\ees
The corresponding energy density has the form
\be\label{rhobnrt}
\rho(r) = \frac{4s S^2}{r^{2D-2}}\,\Bigl(1-2s-(1-3s)S^2\Bigr),
\ee
where we are using $S=\sech\left(2s/((D-2)r^{D-2})\right)$.

		\begin{figure}[t!]
		\centering
		\includegraphics[width=6.2cm,trim={0cm 0cm 0 0},clip]{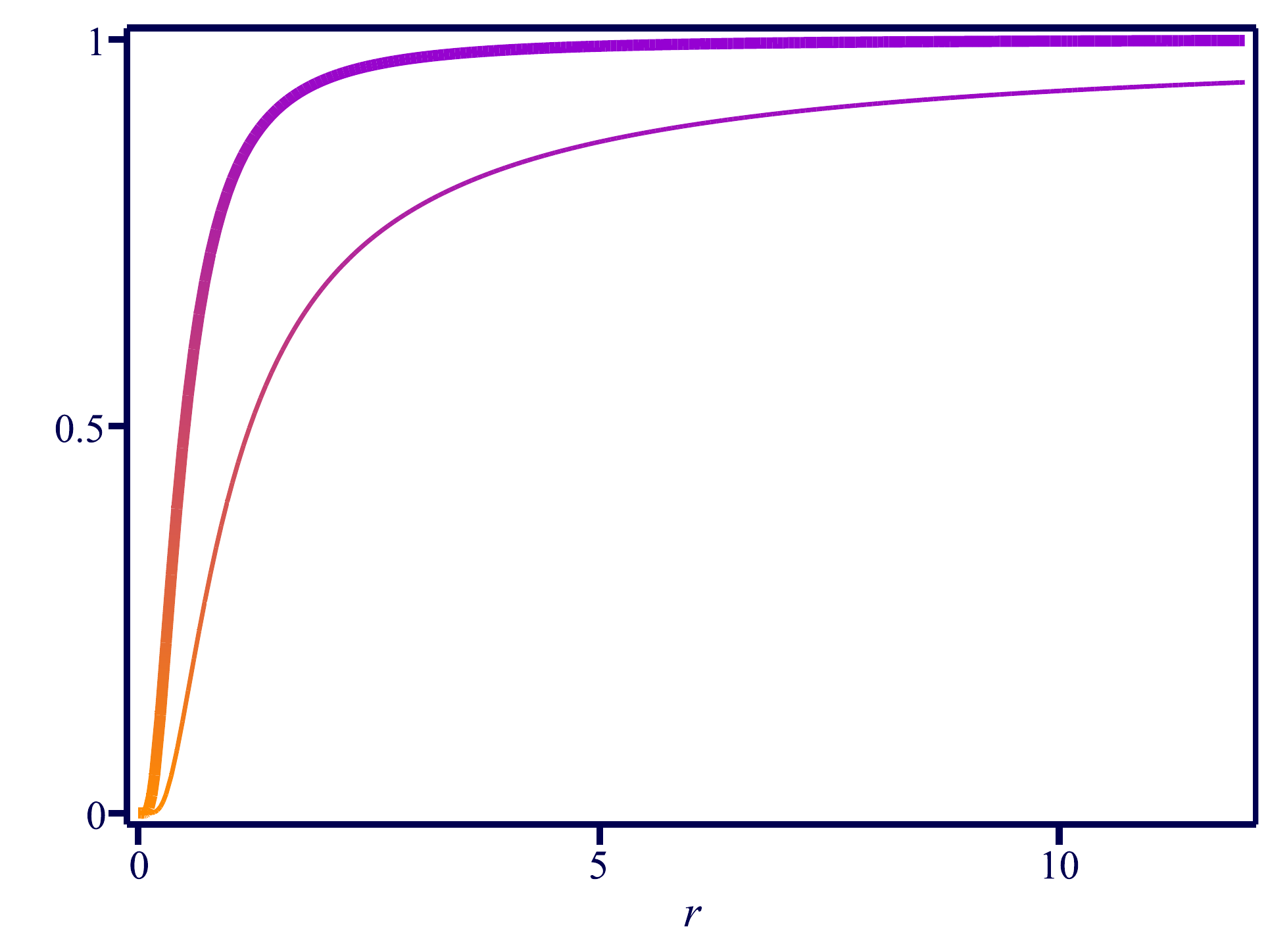}
		\includegraphics[width=6.2cm,trim={0cm 0cm 0 0},clip]{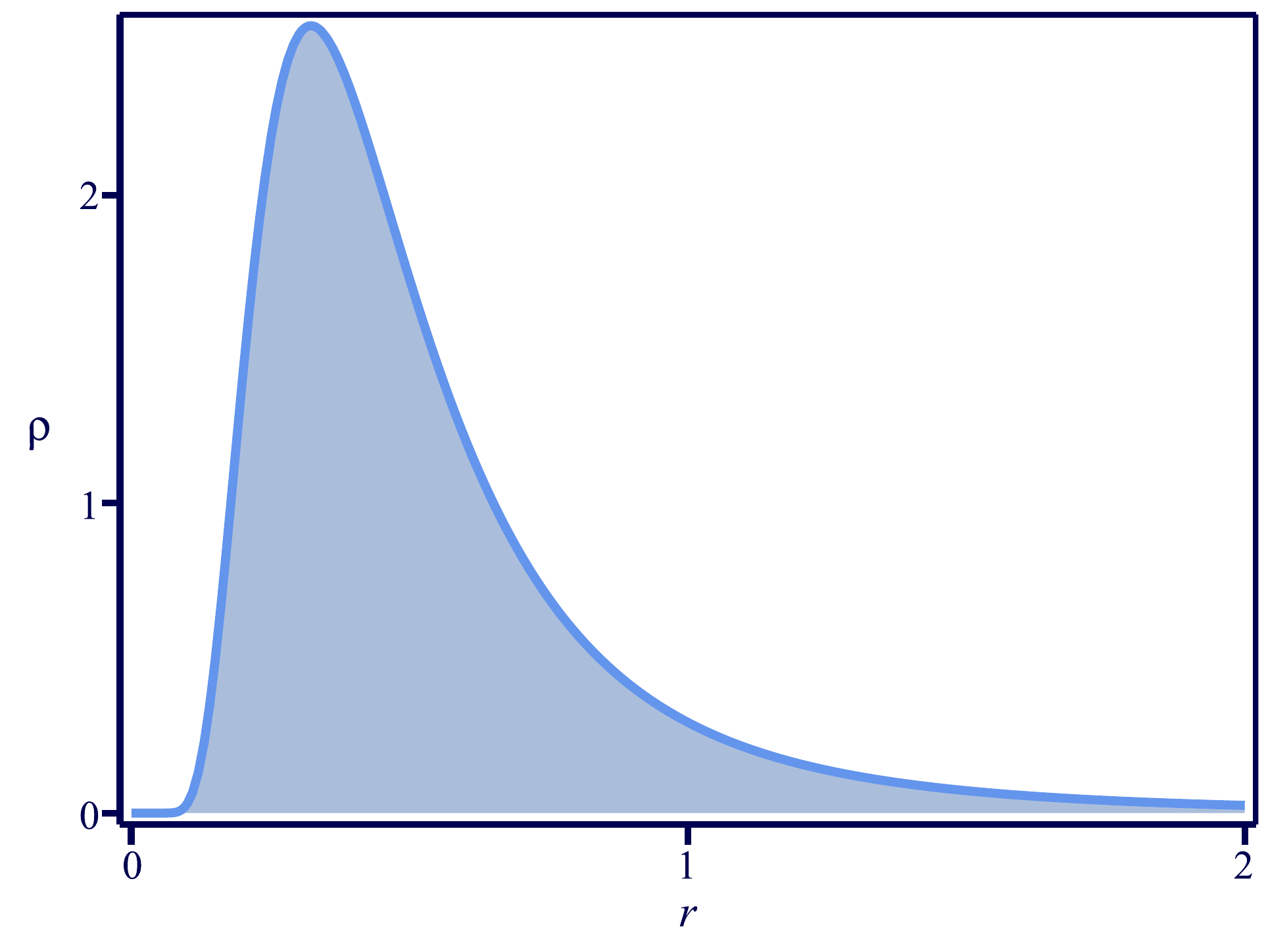}
		\caption{The solutions $H(r)$ and $\Hc(r)$ in Eq.~\eqref{solbnrt} (left), and the energy density \eqref{rhobnrt} (right), for $D=3$ and $s=1/3$. In the left panel, the thinner line represents $H(r)$ and the thicker one stands for $\Hc(r)$.}
		\label{fig3}
		\end{figure}

The above calculations that led us to the analytical solutions are simpler to understand if one changes the $r$ variable to $x$, such that $r^{D-1}({dH}/{dr})={dH}/{dx}$ and $r^{D-1}({d\Hc}/{dr})={d\Hc}/{dx}$. In this case, the first order equations \eqref{foeq} become ${dH}/{dx} = H(2-H)-s\Hc^2 \;\text{and}\; {d\Hc}/{dx} = -2s(H-1)\Hc.$
First order equations similar to these ones were first investigated in \cite{DB1} and then in \cite{DB2,Gui}.

\begin{figure}[t!]
		\centering
		\includegraphics[width=5.0cm]{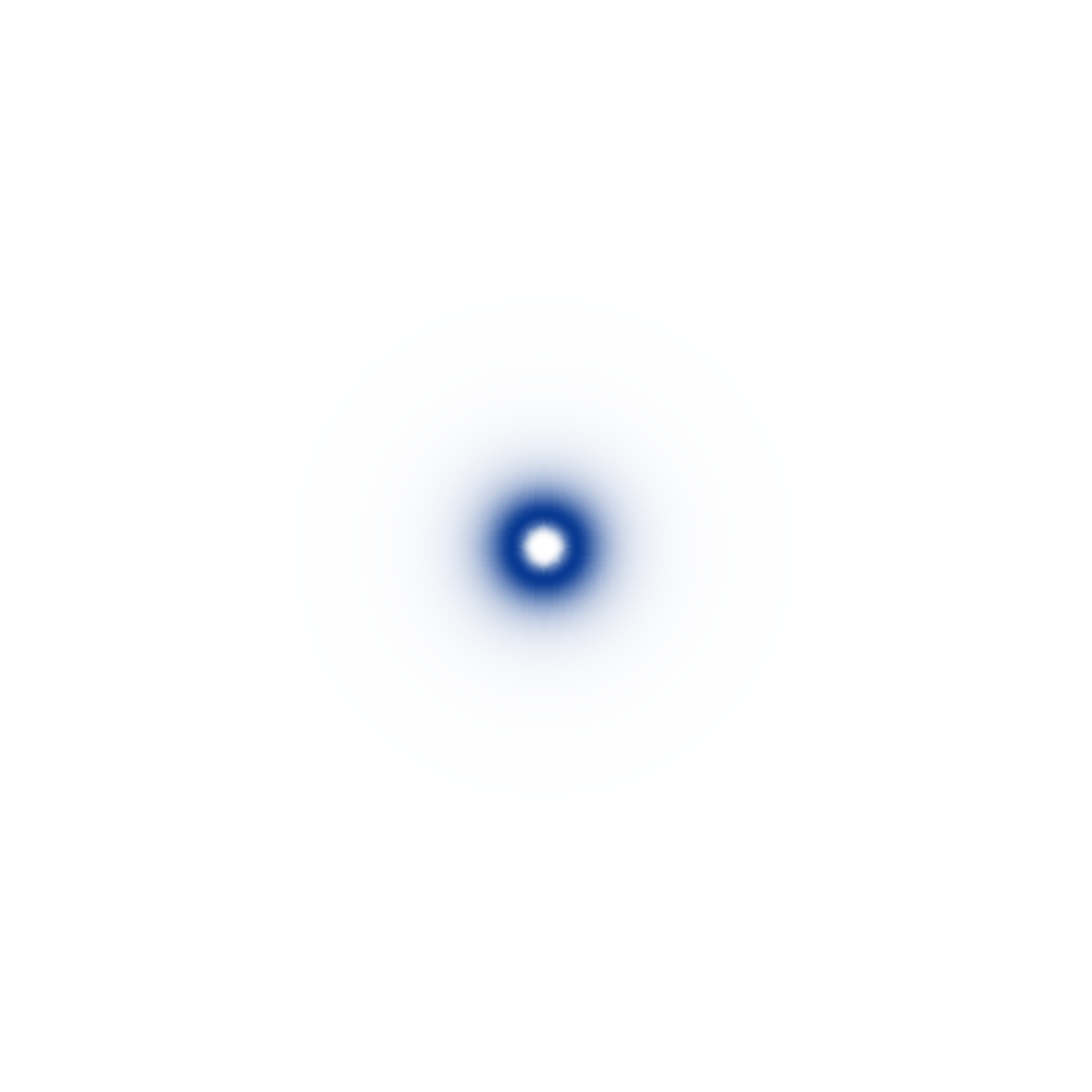}
		\includegraphics[width=5.0cm]{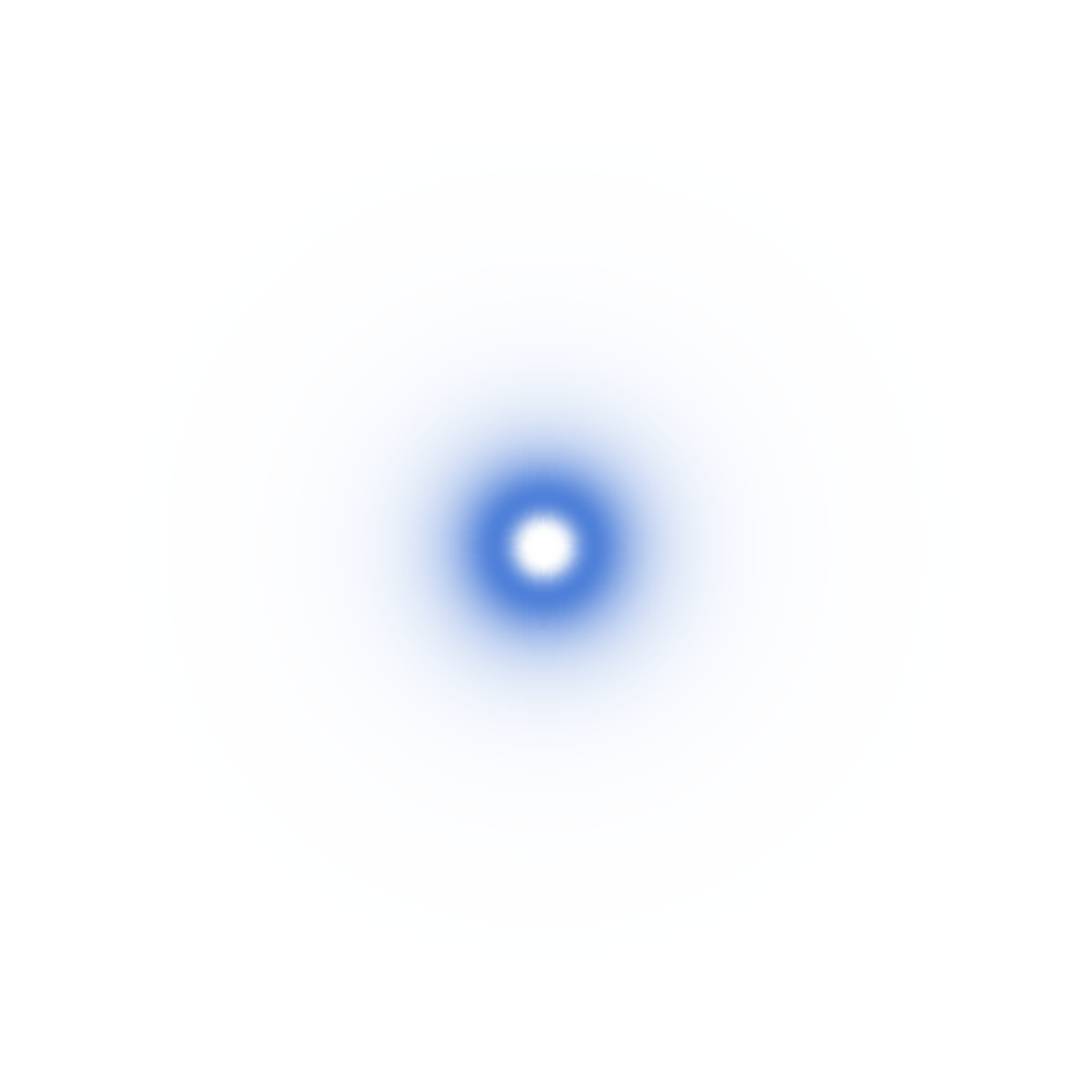}
		\includegraphics[width=5.0cm]{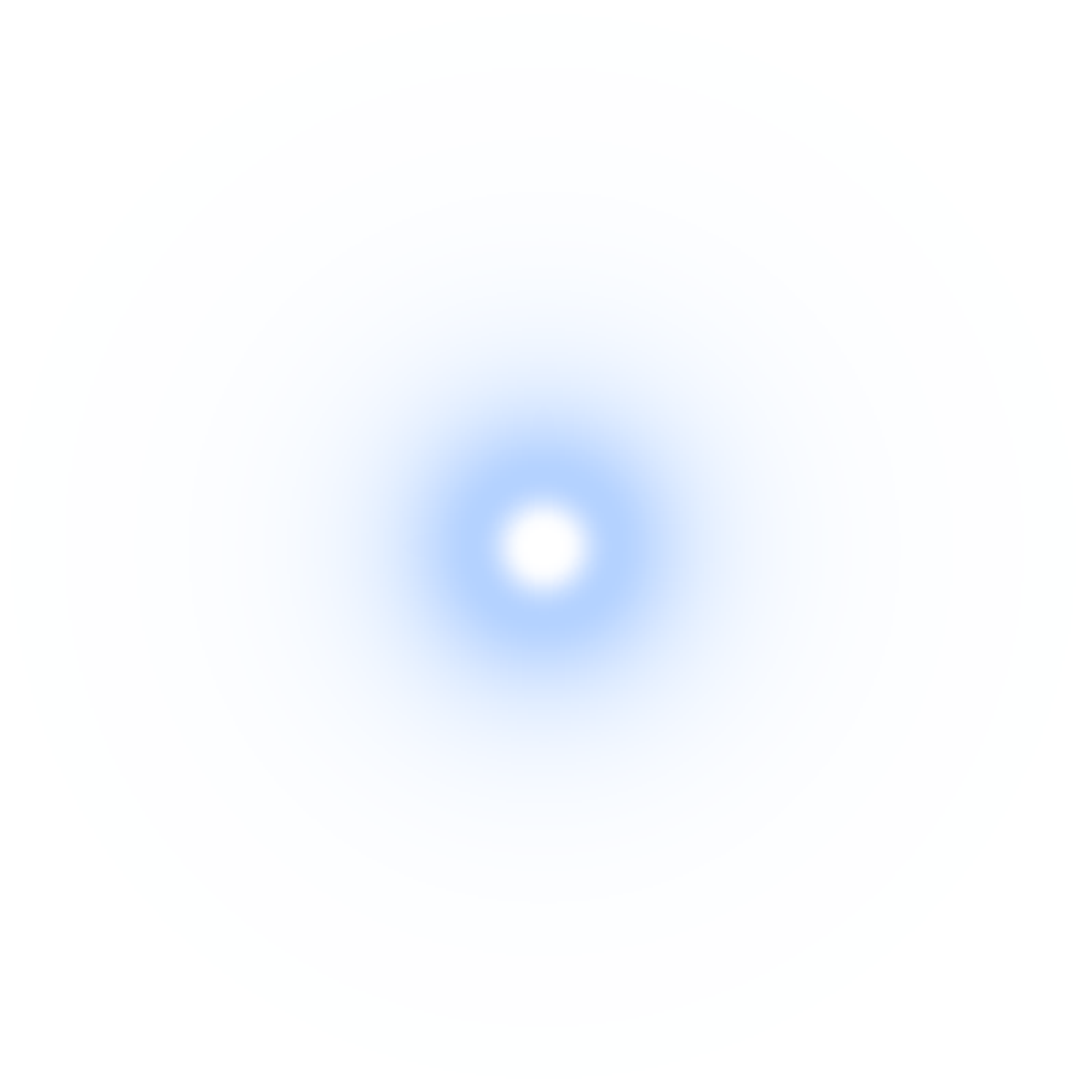}
		\caption{The planar section of the energy density \eqref{rhobnrt} passing through its center, for $D=3$ and $s=0.2$ (left), $s=0.3$ (center), $s=0.4$ (right).}
		\label{fig4}
		\end{figure}

The minimum energy can be obtained from Eq.~\eqref{EB2}: it gives $ E=2\,\Omega_{(D)}/3$, which does not depend on $s$. This is of interest because we can control the Higgs portal without changing the total energy of the localized structure. We depict in Fig.~\ref{fig3} the solutions and energy density for $D=3$ and $s=1/3$. We see that the maximum of the energy density is now around the value $3.0$, much higher than it appeared in the first model, depicted in Fig \ref{fig1}. And also, its hollow portion is now smaller than in the first model. If we compare the first and third models, we see that the addition of the second set of fields contributes to make the monopole smaller but more concentrated around its center. Moreover, we display in Fig. \ref{fig4} the planar section of the energy density for $s=0.2, 0.3$, and $0.4$, to show how the variation of $s$, which describes the interaction between the two sets of scalar fields, may directly contribute to control the profile of the global monopoles. If one compares this with the localized structure displayed in Fig. \ref{fig1}, one realizes the significant effectiveness of the parameter $s$ to control the distribution of matter inside the structure. The extended model then clearly shows the importance of enhancing the symmetry, duplicating the degrees of freedom as an effective manner to control the profile of the solution.

\subsection{Second extended model}

We propose a distinct model, extended to include a single scalar field, $\chi$, but we now include another coupling. The Lagrange density has the form
\be
\LL_3= -\frac12f(\chi) \partial_\mu\phi^a\partial^\mu\phi^a -\frac12 \partial_\mu\chi\partial^\mu\chi-V(|\phi|,\chi).
\ee
Here, $\chi$ is a real scalar field that can appear in the potential, but it also couples with the derivative of the other scalar via the real and non-negative function $f(\chi)$. This coupling is inspired by the recent work \cite{liao}, where it was used to simulate geometric constriction on kinks in the real line. In the present model, the equations of motion are
\bes
\bal
&\partial_{\mu}(f\partial^{\mu}\phi^{a})=\frac{\phi^a}{|\phi|}V_{|\phi|},\\
&\partial_{\mu}\partial^{\mu}\chi=\frac{1}{2}f_{\chi}\partial_{\mu}\phi^{a}\partial^{\mu}\phi^{a}+V_{\chi}.
\eal
\ees
By considering the ansatz \eqref{ansatz} and $\chi=\chi(r)$ as static configurations, we get
\bes
\bal
\frac{1}{r^{D-1}}\left(fr^{D-1} H^\prime \right)^\prime &= \left(D-1\right)\frac{fH}{r^2} + V_{H},\\
\frac{1}{r^{D-1}}\left(r^{D-1} \chi^\prime \right)^\prime &= \frac12f_\chi\left({H^\prime}^2+\left(D-1\right)\frac{H^2}{r^2}\right)+ V_{\chi},
\eal
\ees

\begin{figure}[t!]
		\centering
		\includegraphics[width=6.2cm,trim={0cm 0cm 0 0},clip]{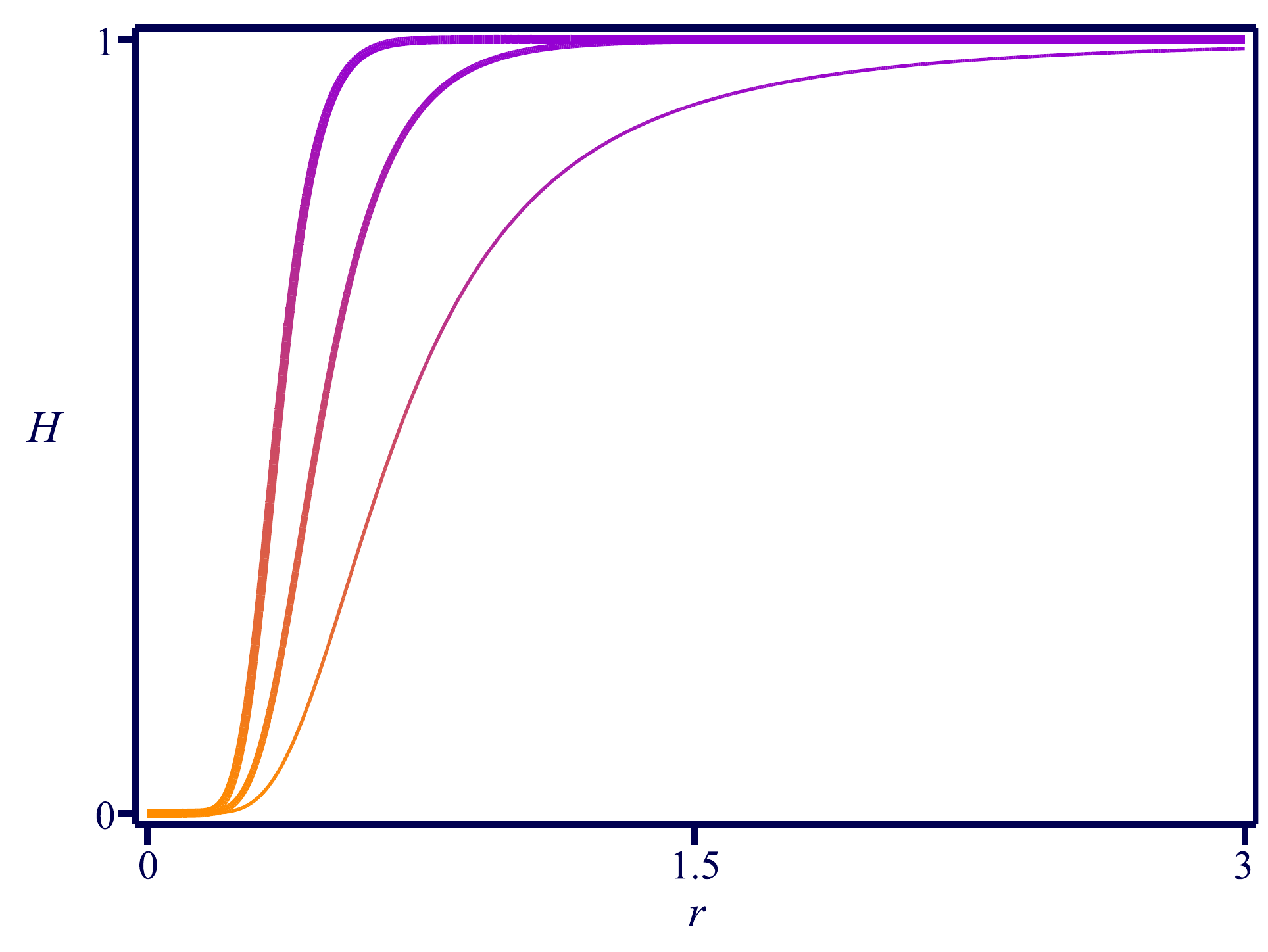}
		\includegraphics[width=6.2cm,trim={0cm 0cm 0 0},clip]{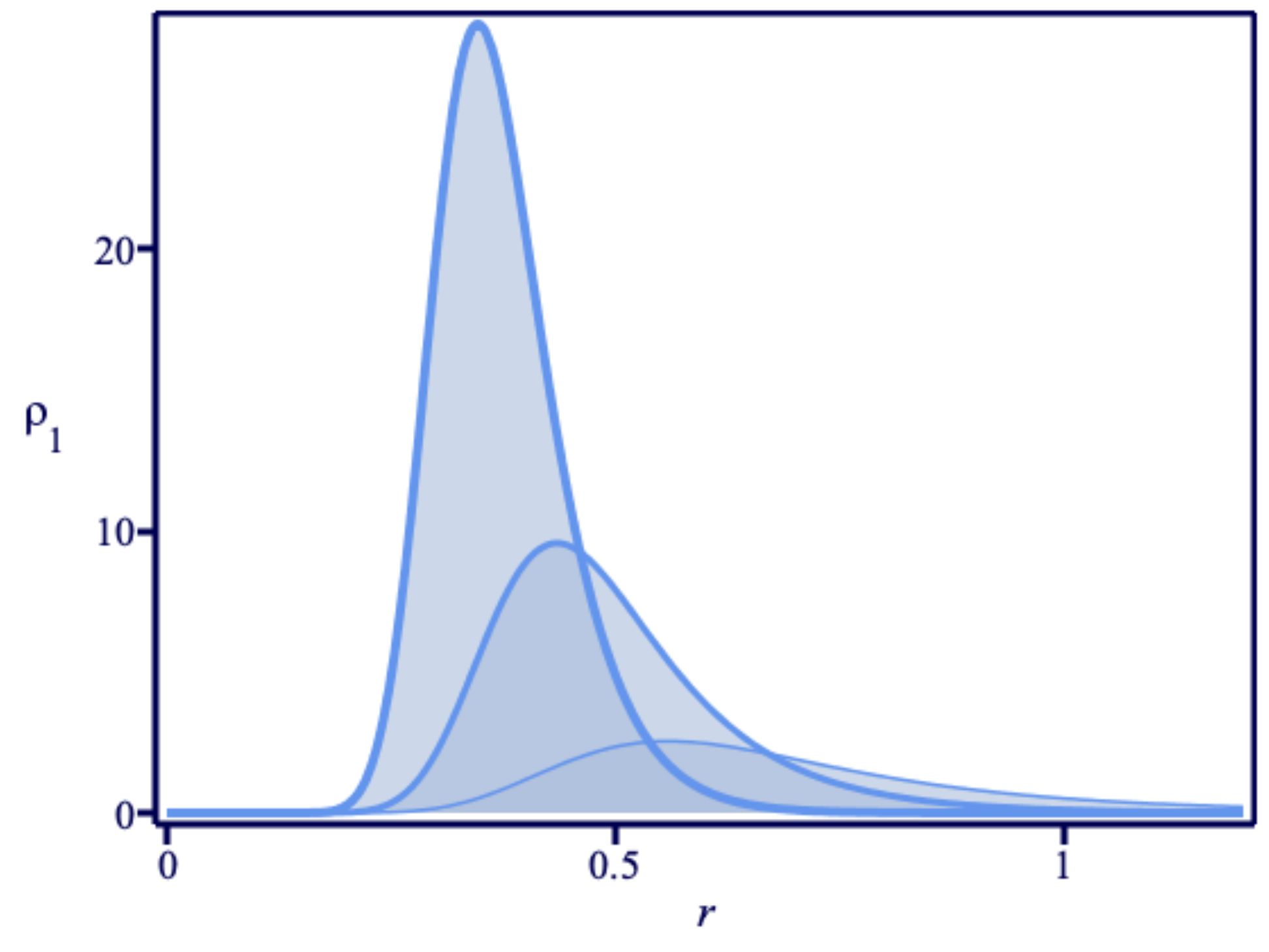}
		\caption{The solution $H(r)$ (left) in Eq.~\eqref{sola} its associated energy density \eqref{rho1} (right) for $D=3$ and $n=1,4$ and $16$. The darkness of the color increases with $n$.}
		\label{fig5}
		\end{figure}

\begin{figure}[t!]
		\centering
		\includegraphics[width=5cm,trim={0cm 0cm 0 0},clip]{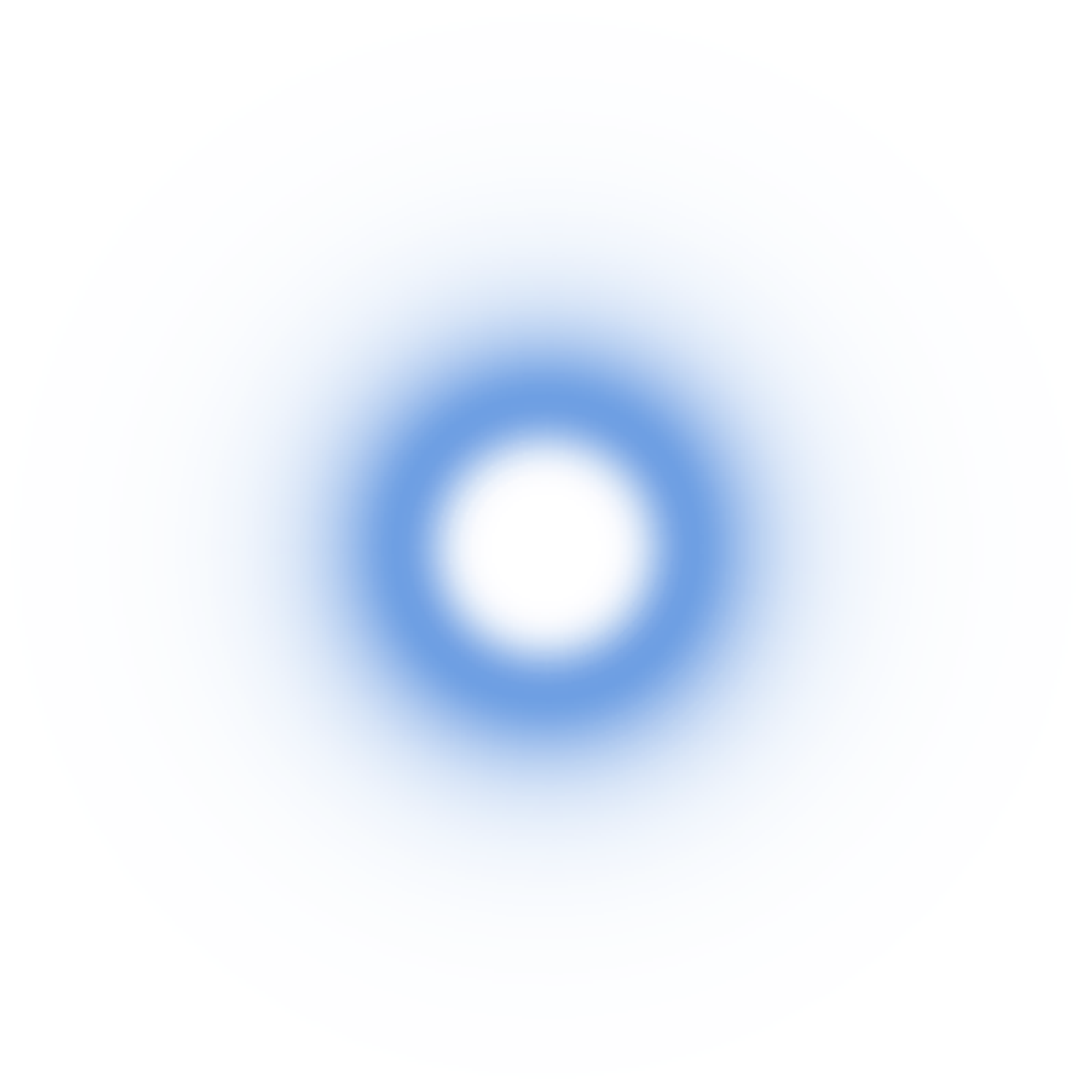}
		\includegraphics[width=5cm,trim={0cm 0cm 0 0},clip]{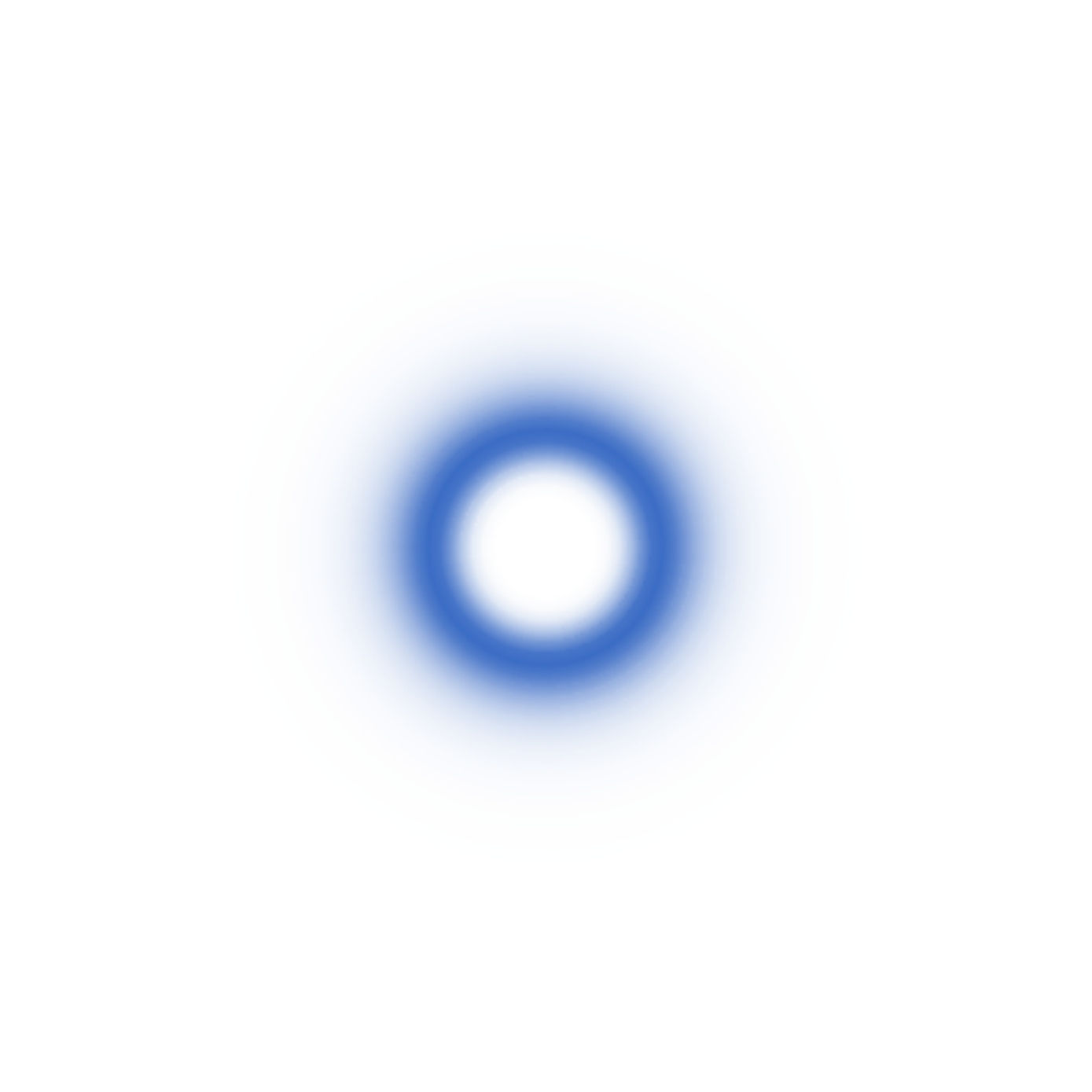}
		\includegraphics[width=5cm,trim={0cm 0cm 0 0},clip]{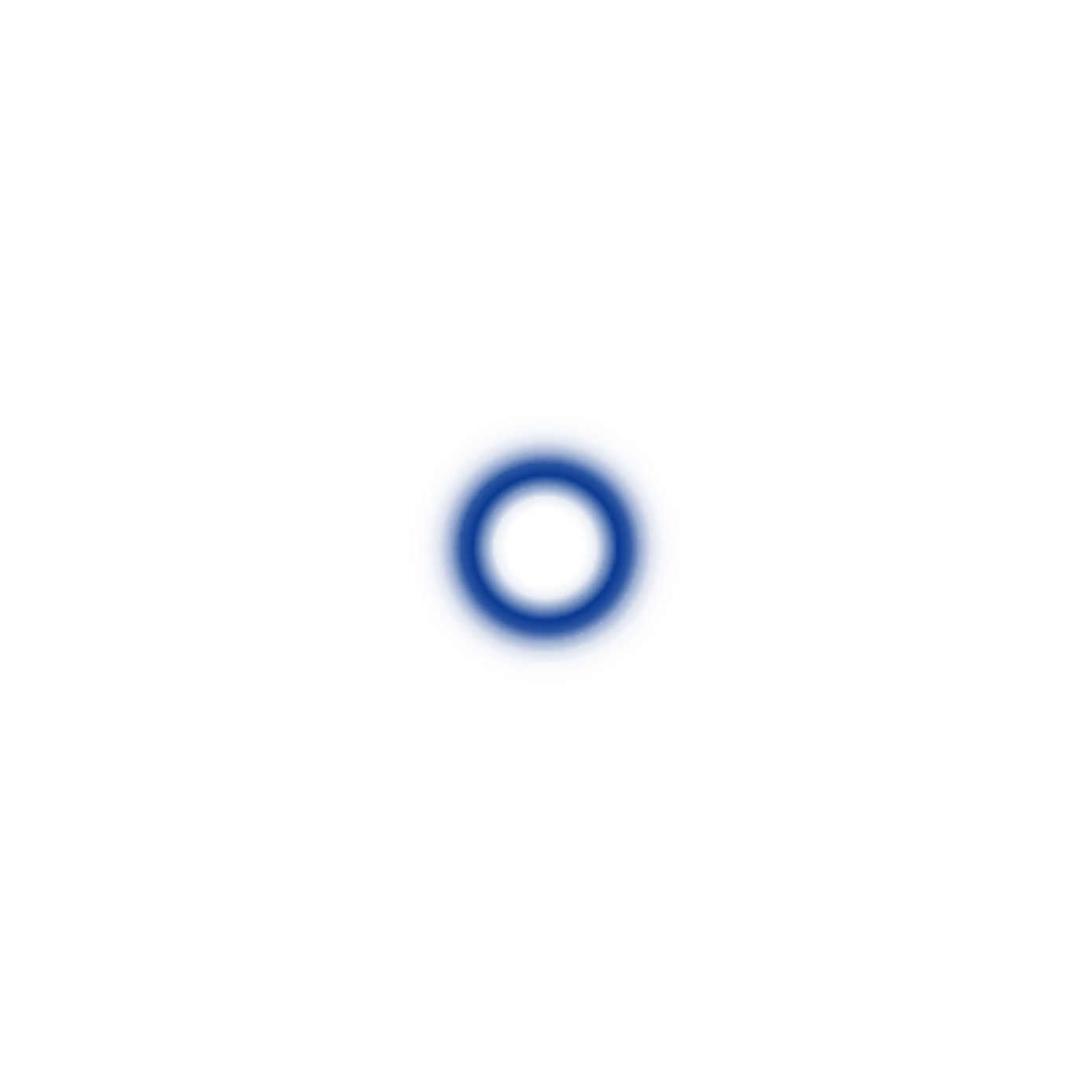}
		\caption{The planar section passing through the center of the energy density \eqref{rho1} for $D=3$ and $n=1$ (left), $4$ (center) and $16$ (right). The darkness of the color increases with $n$}
		\label{fig6}
		\end{figure}

In this case, by following the Bogomol'nyi procedure, the model attains minimum energy, $E=\Omega_{(D)}\,|W(H(\infty),\chi(\infty)) - W(H(0),\chi(0))|$ if the potential has the form
\be
	\label{potentialnew}
V(r,|\phi|,\chi) = \frac{1}{2r^{2D-2}}\left(\frac{W_{|\phi|}^2}{f(\chi)} + W_{\chi}^2\right)-\frac{D-1}{2r^2}f(\chi)|\phi|^2,
\ee
and the following first order equations are obeyed
\be\label{fof}
H^\prime = \frac{W_{H}}{r^{D-1}f(\chi)} \quad\text{and}\quad \chi^\prime = \frac{W_{\chi}}{r^{D-1}}.
\ee
We then take
\be
W(|\phi|,\chi)=|\phi|^2-\frac{1}{3}|\phi|^{3} + \chi^2-\frac{1}{3}\chi^{3}.
\ee
and $f(\chi)=\left(1-\chi\right)^{-2n}$, with $n\in\mathbb{N}^*$. The first order equations \eqref{fof} become
\be\label{fofex}
H^\prime =\frac{(1-\chi)^{2n}H\left(2-H\right)}{r^{D-1}} \quad\text{and}\quad \chi^\prime = \frac{\chi(2-\chi)}{r^{D-1}}.
\ee
Notice that the equation for $\chi$ can be solved independently. In this special situation, in which fields are not coupled in the function $W$, the energy density can be written as two distinct contributions, in the form $\rho=\rho_1+\rho_2$, where
\be\label{rhof}
\rho_1 =\frac{(1-\chi)^{2n}H^2\left(2-H\right)^2}{r^{2D-2}} \quad\text{and}\quad \rho_2=\frac{\chi^2\left(2-\chi\right)^2}{r^{2D-2}}.
\ee
By solving \eqref{fofex}, one gets the solutions
\bes
\bal \label{sola}
H(r)&=1-\tanh\xi,\\ \label{solb}
\chi(r) &= 1-\tanh\bigg(\frac{1}{(D-2)\,r^{D-2}}\bigg),
\eal
\ees
where
\be
\xi=\frac{1}{(D\!-\!2)\,r^{D-2}} - \!\sum_{k=1}^{n}\frac{1}{2k\!-\!1}\tanh^{2k-1}\!\bigg(\!\frac{1}{(D\!-\!2)\,r^{D-2}}\!\bigg).
\ee
The contributions to the energy density are, using Eq. \eqref{rhof}
\bes
\bal\label{rho11}
\rho_1 &= \frac{1}{r^{2D-2}}\,\tanh^{2n}\bigg(\frac{1}{(D-2)\,r^{D-2}}\bigg)\,\sech^4\xi,\\
\rho_2 &= \frac{1}{r^{2D-2}}\,\sech^4\bigg(\frac{1}{(D-2)\,r^{D-2}}\bigg).
\eal
\ees
By integrating these contributions, we get the total energy $E=E_1+E_2$, where $E_1=E_2=2\Omega_{(D)}/3$, matching with the value expected from the expression above Eq.~\eqref{potentialnew}. The profile of the solution \eqref{solb} and its associated energy density, $\rho_2$, given above, can be seen in Fig.~\ref{fig1}, so we do not display it again. The solution \eqref{sola} and its associated energy density \eqref{rho1} is depicted in Fig.~\ref{fig5}. As we can then see, the tail of the structure shrinks as $n$ increases. To illustrate this feature better, we display the planar section of the energy density passing through the center in Fig.~\ref{fig6}. We see that the effect of the shrinking of the tail of the configuration is significant, contributing to form a shell-like structure.


\section{Presence of electric charge}
\label{electric}

Motivated by the above results and the recent investigation on electrically charged localized structures \cite{elect}, let us now consider another model. Here we describe a single point charge in a medium whose electric permittivity is controlled by scalar fields. The model is described by the Lagrange density
\be\label{lcharge}
\mathcal{L}_4=-\frac{1}{2}\partial_\mu\phi^a\partial^\mu\phi^a-\frac{1}{4}P(|\phi|)F_{\mu\nu}F^{\mu\nu} -A_{\mu}j^{\mu},
\ee
where $A^{\mu}$ is the gauge field, $\phi^a$ a scalar field with $D$ components, $F^{\mu\nu}$ is the electromagnetic strength tensor, $P(|\phi|)$ is a dielectric function that  describes a generalized electric permittivity and $j^{\mu}$ is a $(D+1)$-dimensional current that represents an external source. This model is inspired by the recent study \cite{elect} and, compared with the previous model \eqref{model}, it is more involved, including an Abelian gauge field and external current. We consider dimensionless fields and coordinates, and the equations of motion are  
\bes
\begin{align}
    \label{17eq}\partial_{\mu}\partial^{\mu}\phi^{a}-\frac{\phi^{a}}{4|\phi|}P_{|\phi|}F_{\mu\nu}F^{\mu\nu}&=0,\\
   \label{18eq} \partial_{\mu}(P(|\phi|)F^{\mu\nu})-j^{\nu}&=0.
\end{align}
\ees
We take the case of a single point charge at the origin in the absence of currents, in which $j^0= e\delta(r)/r^{D-1}$ and $j^i=0$. In this situation, there is no magnetic field, and the electric field is then described by $E^i= F^{i0}$. By using the ansatz in Eq.~\eqref{ansatz}, one can show that the Gauss' law that arises from Eq.~\eqref{18eq} with $\nu=0$ is solved by
\be
\label{eq24}
\textbf{E}=\frac{e}{r^{D-1}P(H)}\hat{r}.
\ee
Here, $\hat{r}$ denotes the unit vector in the radial direction. Notice that the above electric field depends only on $r$, as expected from the geometrical configuration of the system. The equation of motion \eqref{17eq} for the scalar field with the ansatz in Eq.~\eqref{ansatz} becomes
\be
    \frac{1}{r^{D-1}}\left(r^{D-1} H^\prime \right)^\prime - \left(D-1\right)\frac{H}{r^2} +\frac{1}{2}P_{H}|\textbf{E}|^{2}=0,
\ee
where $P_{H}=\partial P/\partial H$. We can combine it with Eq.~\eqref{eq24} to get
\be
     \frac{1}{r^{D-1}}\left(r^{D-1} H^\prime \right)^\prime = \frac{1}{2r^{2}}\frac{d}{dH}\bigg[\left(D-1\right)H^{2}+\frac{e^{2}}{r^{2D-4}P}\bigg].
\ee
The energy density associated to the above field configurations can be written in the form
\be\label{rhototal}
    \rho = \frac12{H^\prime}^2 + \frac{D-1}{2}\frac{H^2}{r^2} + \frac{1}{2P}\frac{e^{2}}{r^{2D-2}}.
    \ee
Its corresponding energy can be found by integrating the above expression. At this point, we follow similar steps as in the previous section and introduce the auxiliary function $W(|\phi|)$ to develop the Bogomol'nyi procedure for our model. We then write
\be\begin{aligned}\label{rho2}
    \rho = \frac12\left(H^\prime \mp \frac{W_H}{r^{D-1}} \right)^2 +\frac{1}{2P(H)}\frac{e^{2}}{r^{2D-2}}
    + \frac{D-1}{2}\frac{H^2}{r^2}-\frac12\frac{W_H^2}{r^{2D-2}}\pm \frac{1}{r^{D-1}}W^\prime.
\end{aligned}\ee
Since $H=|\phi|$, we now suppose that the dielectric function has the form
\be\label{die}
P=e^{2}\bigg(W^{2}_{|\phi|}-(D-1)|\phi|^{2}r^{2D-4}\bigg)^{-1},
\ee
which makes the second, third and fourth terms in the above expression \eqref{rho2} add to zero. In this case, the energy is bounded from bellow, $E\geq E_B$, with $E_B$ given in the same form of Eq.~\eqref{EB}. Thus, $E$ is minimized to $E=E_B$ for solutions that obey the first order equation \eqref{firstorder}. We then see that the field configurations which arise from the model \eqref{model} can be used to model electrical properties of the medium in which the charge is immersed. Notice that, even though the first order equations involved are similar to the ones in Ref.~\cite{elect}, the above equation presents a negative contribution due to the angular contribution of the gradients of the fields, as they appear in \eqref{ansatz} and this modify the behavior of the electric fields as we shall see bellow. 
The similarity of the mathematical structure of the above model and the global monopole investigated in the previous Sec. \ref{monopole} allows that we think of the inclusion of the electric charge as a natural way of breaking translation invariance in the present model.

		\begin{figure}[t!]
		\centering
		\includegraphics[width=6.2cm,trim={0cm 0cm 0 0},clip]{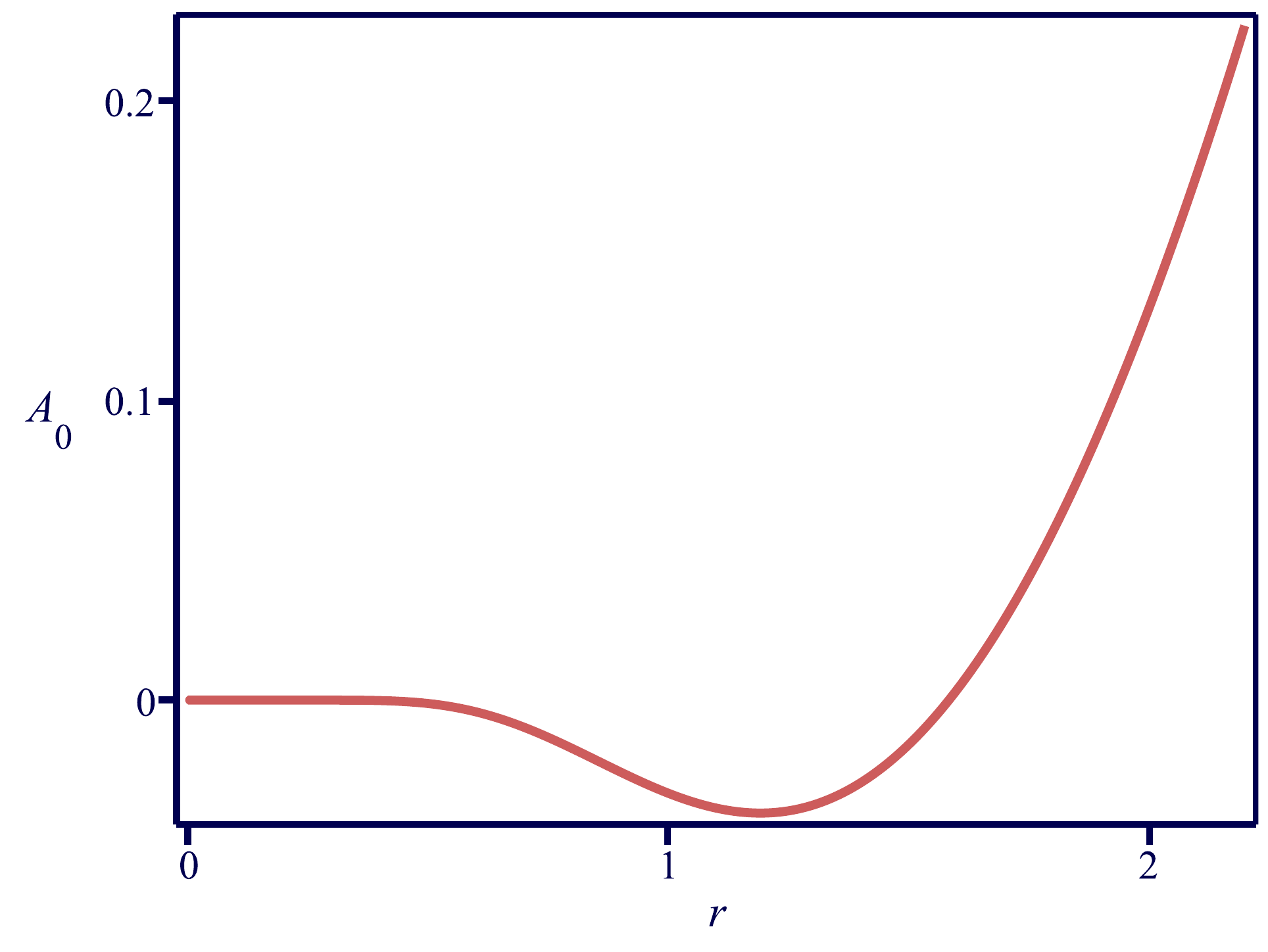}
		\includegraphics[width=6.2cm,trim={0cm 0cm 0 0},clip]{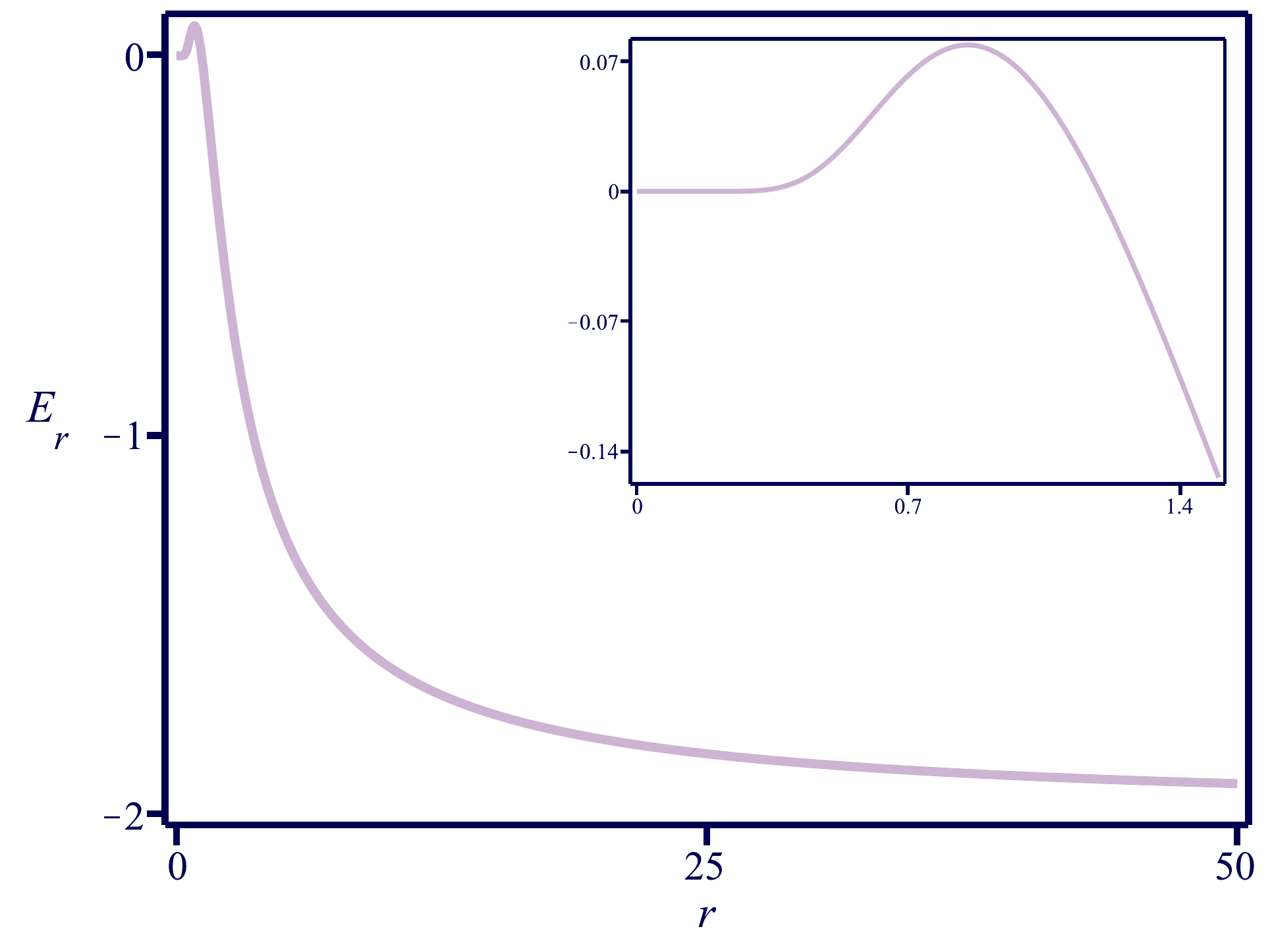}
		\caption{The gauge field (left) and the radial component of the electric field in Eq.~\eqref{ephi4} (right) associated to the system with electric permittivity in Eq.~\eqref{perm1}, for $D=3$ and $e=1$.}
		\label{fig7}
		\end{figure}
		\begin{figure}[t!]
		\centering
		\includegraphics[width=5cm,trim={0cm 0cm 0 0},clip]{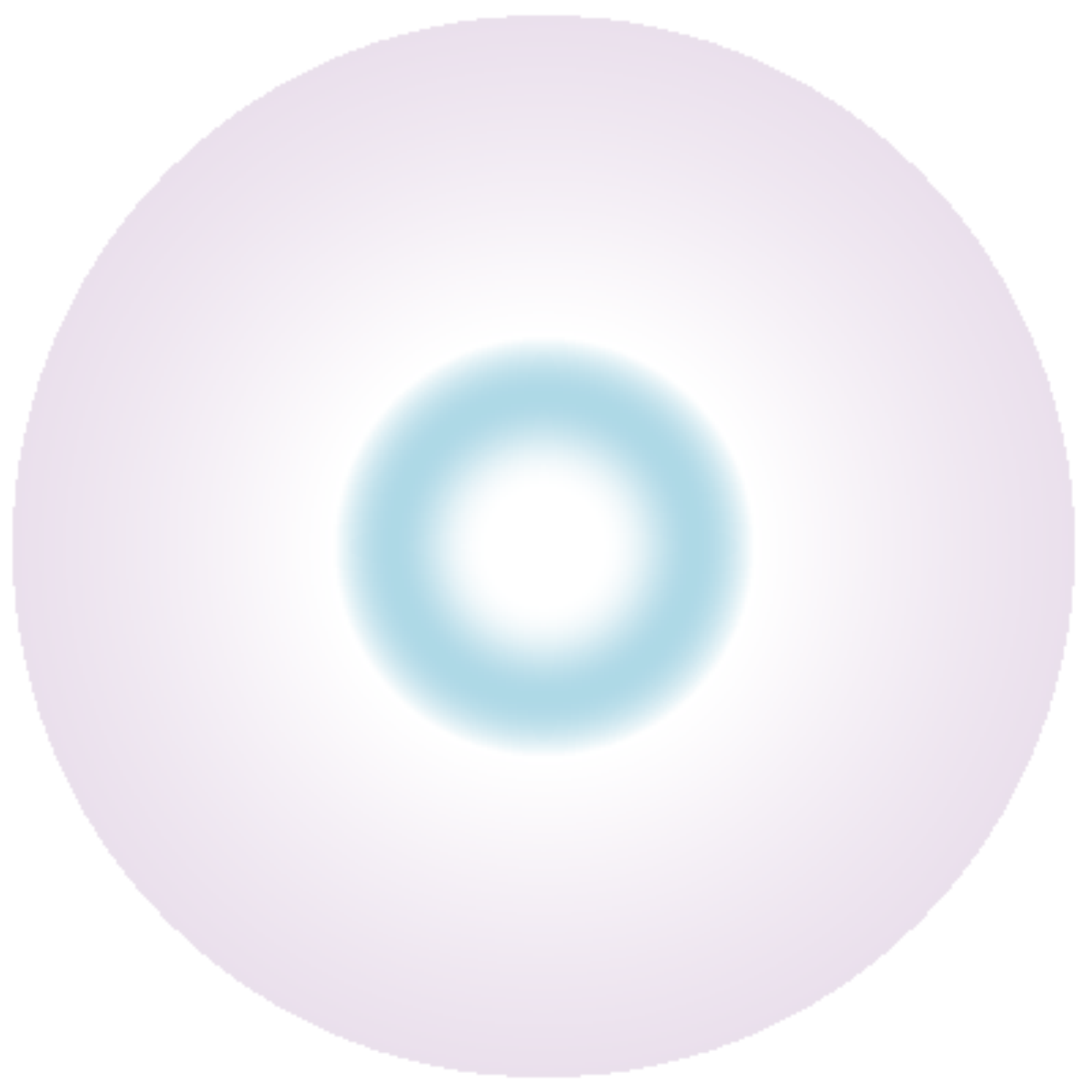}
		\caption{The section of the electric field passing through the center of the structure in the interval $r\in[0,3]$ for $D=3$ and $e=1$.  White, blue and purple are used to represent $E_r=0$, $E_r=0.079$ and $E_r=-2$, respectively..}
		\label{fig8}
		\end{figure}

To illustrate the above results, we take the model described by the function in Eq.~\eqref{p1}, introduced to control the $|\phi|^4$ model investigated in the previous Sec. \ref{monopolea}. In this situation, the dielectric function \eqref{die} is given by
\be\label{perm1}
    P=e^{2}\left(|\phi|^2\left(2-|\phi|\right)^2-(D-1)r^{2D-4}|\phi|^2\right)^{-1}
\ee
and the electric field in Eq.~\eqref{eq24} has the form
\be\label{ephi4}\begin{aligned}
\textbf{E}=\frac{1}{e}\Bigg[\frac{1}{r^{D-1}}\sech^{4}{\bigg(\frac{1}{(D-2)r^{D-2}}\bigg)}
   -\left(1-\tanh{\bigg(\frac{1}{(D-2)r^{D-2}}\bigg)}\right)^{2}(D-1)r^{D-3}\Bigg]\hat{r} \;.
\end{aligned}
\ee
We consider $\textbf{E}=E_r(r)\hat{r}$ and use the above expression to calculate the gauge field with the equation $dA_0/dr=-E_r$. We take $A_0(0)=0$, for simplicity, and display both $A_0(r)$ and $E_r(r)$ in Fig.~\ref{fig7} for $D=3$. Also, in Fig. \ref{fig8} we depict the spatial section of the electric field passing through the center of the structure, with the color purple representing $E_r=-2$, white for $E_r=0$ and blue for the maximum, $E_r\approx0.079$. Notice that the electric field generated by a positive charge is null at the origin, then is positive as $r$ increases and then becomes negative outside the radius $r\approx1.192$, pointing towards the center. This is the opposite behavior of the usual Coulomb's law since it attracts positive charges, and this remind us of the similar effect that appeared before for vortices in planar systems, in the case of magnetic flux inversion in fractional vortices in the presence of two-component superconductors \cite{vortice,nature}. For any dimensions, $D\geq3$, the electric field at the origin is null, $\textbf{E}(0)=0$. The energy density in Eq.~\eqref{rhototal} is such that $\rho$ has the same form of Eq.~\eqref{rho1}. The energy is calculated with Eq.~\eqref{EB}, which leads us to $E=2\Omega_{(D)}/3$.

\begin{figure}[t!]
		\centering
		\includegraphics[width=6.2cm,trim={0cm 0cm 0 0},clip]{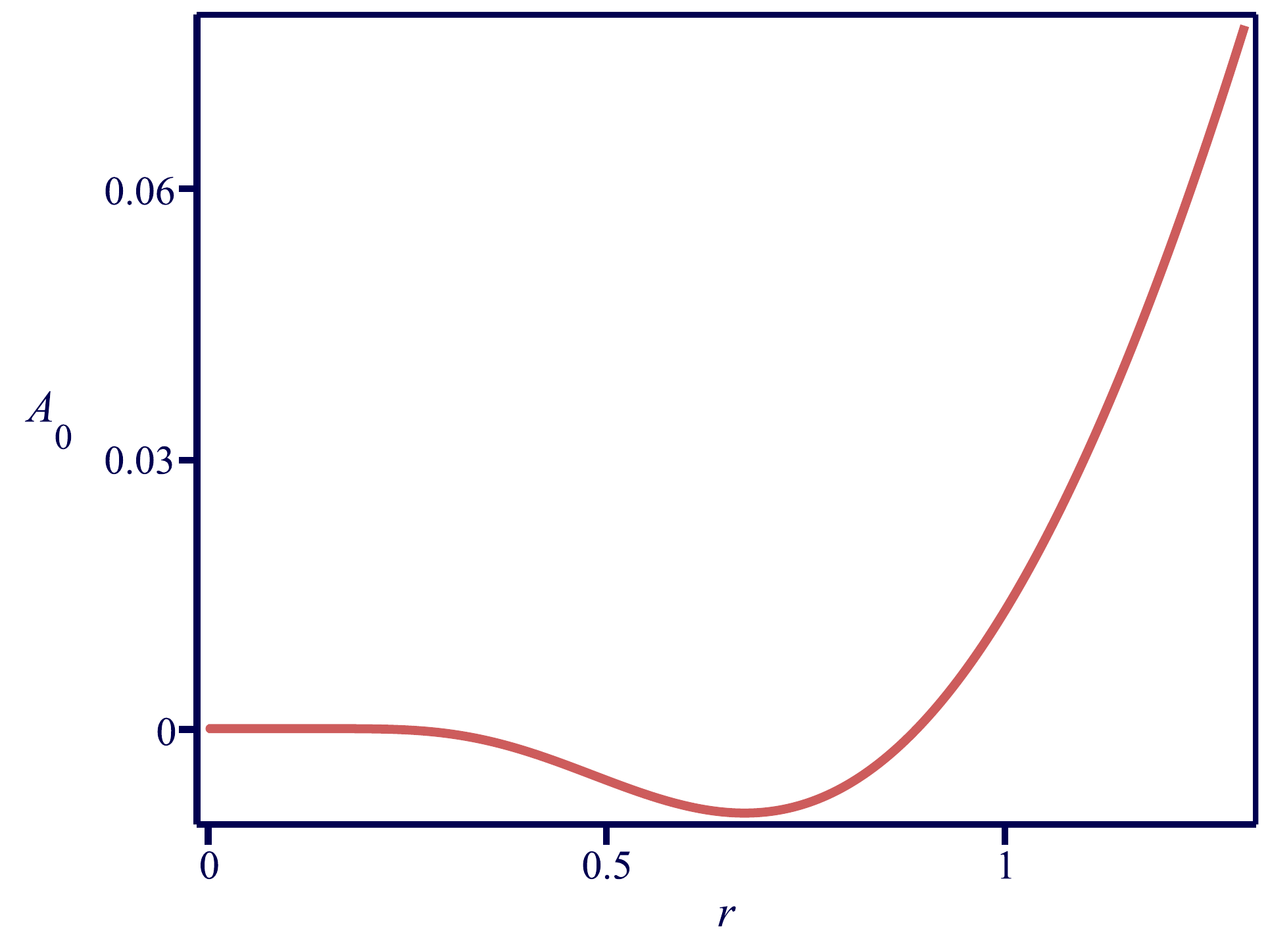}
		\includegraphics[width=6.2cm,trim={0cm 0cm 0 0},clip]{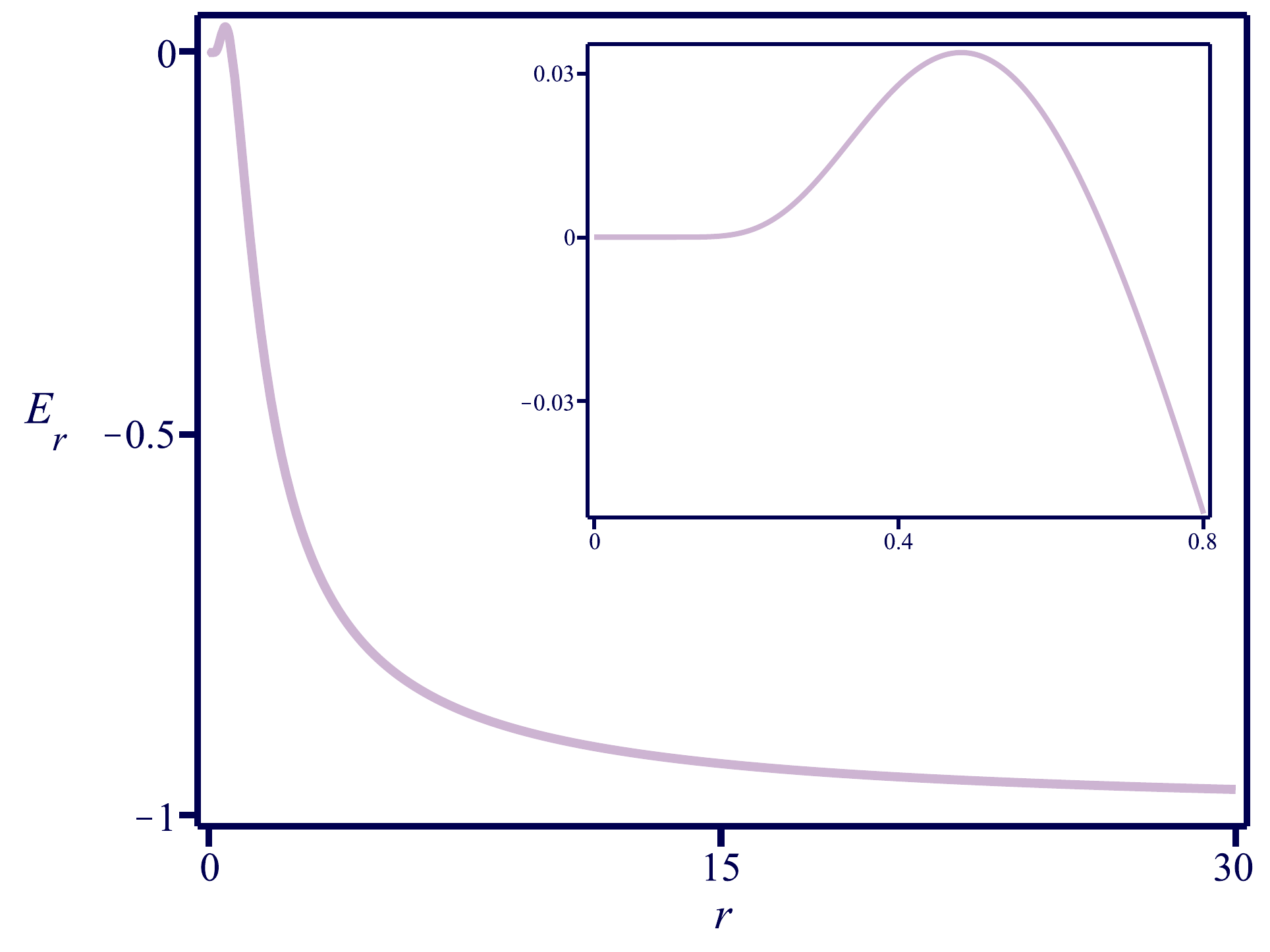}
		\caption{The gauge field (left) and the electric field in Eq.~\eqref{ephi6} (right) associated to the system with electric permittivity in Eq.~\eqref{perm2}, for $D=3$ and $e=1$. The inset in the right panel highlights the behavior of the electric field near the origin.}
		\label{fig9}
		\end{figure}

\begin{figure}[t!]
		\centering
		\includegraphics[width=5cm,trim={0cm 0cm 0 0},clip]{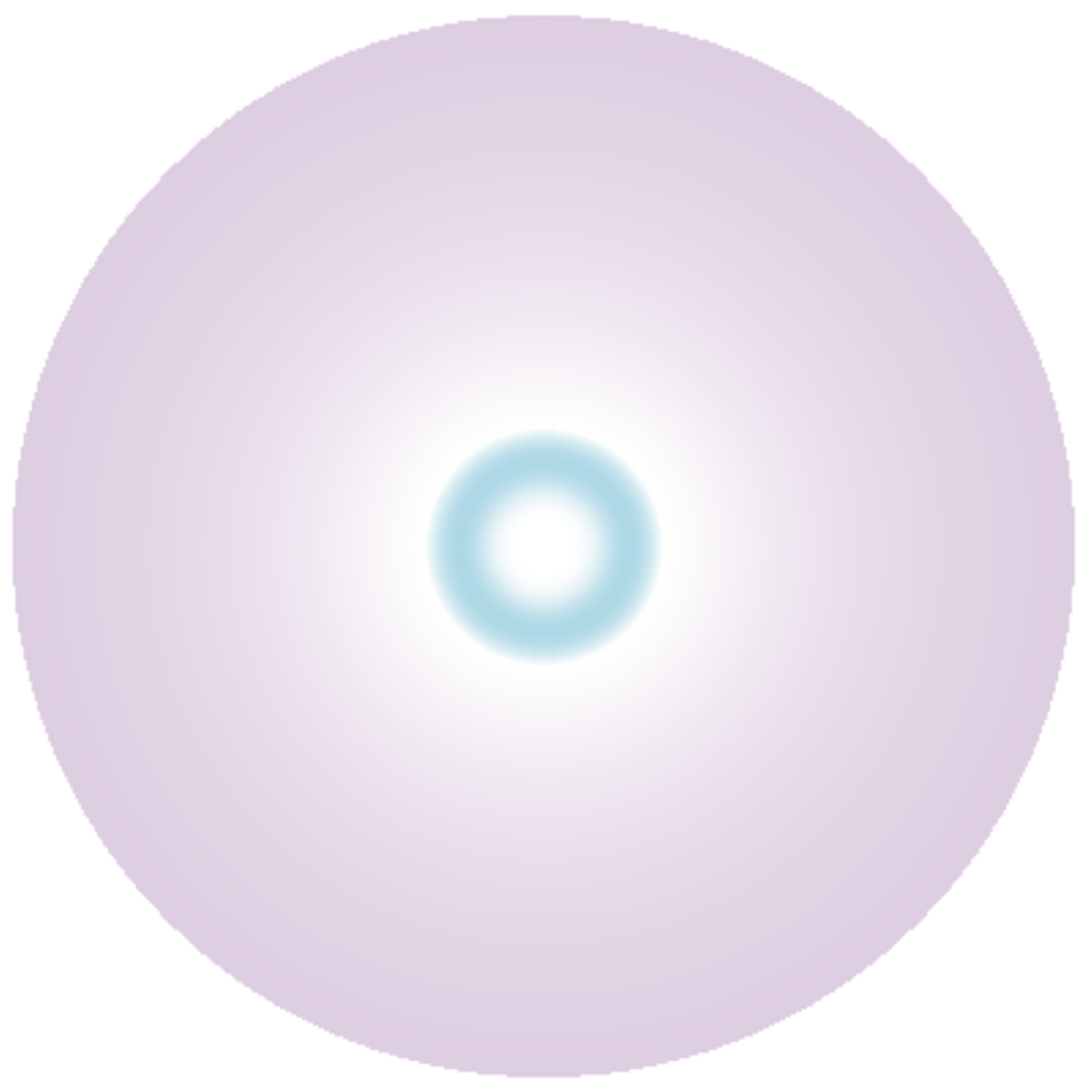}
		\caption{The section of the electric field passing through the center of the structure in the interval $r\in[0,3]$ for $D=3$ and $e=1$. White, blue and purple are used to represent $E_r=0$, $E_r=0.034$ and $E_r=-1$, respectively.}
		\label{fig10}
		\end{figure}

Another example can be considered with the $W$ function displayed in Eq.~\eqref{p2}, which was used to describe the $|\phi|^6$ model investigated in Sec. \ref{monopoleb}. In this case, the dielectric function \eqref{die} becomes
\be\label{perm2}
    P=e^{2}\left(|\phi|^2\left(\left(1-|\phi|^2\right)^2-(D-1)r^{2D-4}\right)\right)^{-1}
\ee
and the electric field in Eq.~\eqref{eq24} is
\be\label{ephi6}\begin{aligned}
\textbf{E}=\frac{1}{8e}\bigg{\{}\frac{1}{r^{D-1}}\bigg[\sech^{2}\!\left(\frac{1}{(D-2)\,r^{D-2}}\right)\!\!+\!4(D\!-\!1)r^{2D-4}\bigg]
    \;\bigg[1 +\tanh\left(\frac{1}{(D-2)\,r^{D-2}}\right)\bigg] -8(D-1)r^{D-3}\bigg{\}}\;\hat{r}.
\end{aligned}
\ee
The electric field at the origin is null, for all $D\geq 3$. We use the above expression to calculate the gauge field, $A_0$, as in the previous example. In Fig.~\ref{fig9}, we display both $A_0$ and $E_r$ for $D=3$ and $e=1$. In Fig.~\ref{fig10}, we display the electric field in the plane. One can see that the electric field starts at zero at the origin, increases until getting a maximum and then decreases crossing the zero and going to negative values until getting to $-1$ at very large radial distances. The energy density is given by Eq.~\eqref{rhophi6}, such that the associated energy in Eq.~\eqref{EB} is $E=3\Omega_{(D)}/16$.

As we have commented before in \cite{elect}, the results on spatially localized structures with electric profile described above may be of interest to ferroelectrics and multiferroics mateirals, since the investigation of topological domains in such materials have attracted interest in the recent years, due to the potential application in the construction of nanoelectronic devices.

\section{Conclusion}
\label{end}

In this paper we introduced a procedure to obtain finite energy global monopoles. By investigating the behavior of the energy density, we calculated the contribution that usually leads to infinite energy and, by developing the Bogomol'nyi procedure, we found that the potential may be written in a specific form to balance the energy. We also showed that this formalism leads us to first order equations compatible with the equations of motion and, as a bonus, the energy can be calculated in terms of the boundary conditions associated to the solutions.

The presence of the first order equations allows for the interesting feature of making a connection with kink solutions in $(1,1)$ spacetime dimensions. One must be careful, however, since the correspondence between these two distinct scenarios works due to the fact that the monopole solution only maps half of the kink configuration.

As we showed in Eq.~\eqref{potential}, the potential engender a negative term that depends explicitly on the radial coordinate. This suggests that we care about the stability of our configurations. For this reason, we investigated the linear stability of the field configuration and showed that the operator that drives the stability can be factorized in terms of supersymmetric partners, in this sense ensuring stability of the solution under small fluctuations. Moreover, in addition to the two $|\phi|^4$ and $|\phi|^6$ models studied in Sec. \ref{monopole}, we have also investigated extended models considered in Sec. \ref{extended}. We first considered a model composed of two sets of scalar fields that interact with one another via the coupling between them, in a way similar to the Higgs portal recently reviewed in \cite{HP}. There we have constructed localized structures that may be small and hollow. We also considered another model, in which we added an extra scalar field to interact with the other field in a way similar to the case investigated before in \cite{liao}, intending to simulate a geometric constriction. As we have shown, the model is capable of generating a localized structure with the energy density shrinking to form a localized shell. We are now investigating other possible extensions, in a way similar to the investigations implemented before concerning small and hollow magnetic monopoles \cite{olmo}, and monopoles with internal bimagnetic composition \cite{BI}.     

Additionally, we have studied the extended model \eqref{lcharge}, which includes the presence of a single point charge in a medium with generalized  electric permittivity. In this scenario, one may also find minimum energy configurations and, as noticed, the first order equations and energy obtained are the very same of the global monopole previously studied. However, the charge generates an electric field whose behavior differs from the usual Coulomb's law, since it may point towards a positive charge, due to the change of its sign along the radial coordinate.

The first order framework which we described in the present work leads to stable configurations and suggests that the model \eqref{model}, for instance, may be the bosonic portion of a supersymmetric theory, so one may investigate supersymmetric extensions of it. Another issue of interest concerns the study of models with enhanced symmetries, in the lines of Refs.~\cite{wit,shi,schaposnik1,schaposnik2,magint}. One interesting possibility is to consider the inclusion of local gauge symmetries; a model of current interest could have the symmetry $SU(2)\times O(3)$, with the local $SU(2)$ symmetry composed with the global $O(3)$ symmetry, for instance. This could allow to investigate how a global monopole could modify the solution of a local monopole. The interest in mixing local and global symmetries goes beyond high energy physics, and can also be used to describe spin ice systems with enhanced symmetries; see, e.g., Refs. \cite{SI1,SI2,SI3} and references therein. Another issue concerns the presence of gravity, which was not considered in this work. For instance, since in some magnetic monopole solutions described in \cite{g2,g3,g6}, gravity may plays an important role in the study of stability, it seems of current interest to investigate the behavior and stability of solutions in the presence of non-trivial geometric backgrounds. This is also directly connected with the recent study \cite{morris}, in which the author develops several distinct investigations that explore stable field configurations in radially symmetric spacetimes, supporting, for instance, spherical bubble, Schwarzschild, cosmic string and whormhole backgrounds. We hope that the present work may foster new research in the subject. 

\acknowledgements{The work is supported by the Brazilian agencies Coordena\c{c}\~ao de Aperfei\c{c}oamento de Pessoal de N\'ivel Superior (CAPES), grant No.~88882.440276/2019-01 (MP), by Conselho Nacional de Desenvolvimento Cient\'ifico e Tecnol\'ogico (CNPq), grants Nos. 404913/2018-0 (DB) and 303469/2019-6 (DB), and by Paraiba State Research Foundation (FAPESQ-PB) grant No. 0015/2019.}

\end{document}